\documentclass[aps,prl,reprint]{revtex4-2}

\usepackage[colorlinks=true,citecolor=blue]{hyperref}
\usepackage{graphicx}
\usepackage{amsmath}
\usepackage{amsfonts}
\usepackage{comment}
\usepackage{verbatim}

\begin{document}

\title{Weak-Memory Dynamics in Discrete Time}

\author{Hugues Meyer and Kay Brandner}

\affiliation{School of Physics and Astronomy, University of Nottingham, Nottingham NG7 2RD, United Kingdom}
\affiliation{Centre for the Mathematics and Theoretical Physics of Quantum Non-Equilibrium Systems, University of Nottingham,
Nottingham NG7 2RD, United Kingdom}

\newcommand{\sA}{\mathsf{A}}
\newcommand{\sB}{\mathsf{B}}
\newcommand{\sC}{\mathsf{C}}
\newcommand{\sD}{\mathsf{D}}
\newcommand{\sE}{\mathsf{E}}
\newcommand{\sF}{\mathsf{F}}
\newcommand{\sG}{\mathsf{G}}
\newcommand{\sH}{\mathsf{H}}
\newcommand{\sI}{\mathsf{I}}
\newcommand{\sJ}{\mathsf{J}}
\newcommand{\sK}{\mathsf{K}}
\newcommand{\sL}{\mathsf{L}}
\newcommand{\sM}{\mathsf{M}}
\newcommand{\sN}{\mathsf{N}}
\newcommand{\sO}{\mathsf{O}}
\newcommand{\sP}{\mathsf{P}}
\newcommand{\sQ}{\mathsf{Q}}
\newcommand{\sR}{\mathsf{R}}
\newcommand{\sS}{\mathsf{S}}
\newcommand{\sT}{\mathsf{T}}
\newcommand{\sU}{\mathsf{U}}
\newcommand{\sV}{\mathsf{V}}
\newcommand{\sW}{\mathsf{W}}
\newcommand{\sX}{\mathsf{X}}
\newcommand{\sY}{\mathsf{Y}}
\newcommand{\sZ}{\mathsf{Z}}

\newcommand{\mD}{\mathcal{D}}
\newcommand{\mE}{\mathcal{E}}
\newcommand{\mF}{\mathcal{F}}
\newcommand{\mG}{\mathcal{G}}
\newcommand{\mH}{\mathcal{H}}
\newcommand{\mI}{\mathcal{I}}
\newcommand{\mK}{\mathcal{K}}
\newcommand{\mQ}{\mathcal{Q}}
\newcommand{\mT}{\mathcal{T}}
\newcommand{\mU}{\mathcal{U}}
\newcommand{\mV}{\mathcal{V}}
\newcommand{\mY}{\mathcal{Y}}
\newcommand{\mZ}{\mathcal{Z}}

\newcommand{\bC}{\mathbb{C}}
\newcommand{\bN}{\mathbb{N}}
\newcommand{\bR}{\mathbb{R}}

\newcommand{\fB}{\mathfrak{B}}
\newcommand{\fT}{\mathfrak{T}}

\newcommand{\trans}{\mathsf{T}}
\newcommand{\tr}{\mathrm{Tr}}

\newcommand{\nrm}[1]{\lVert #1 \rVert}
\newcommand{\abs}[1]{\lvert #1 \rvert}
\newcommand{\ket}[1]{\lvert #1 \rangle}
\newcommand{\bra}[1]{\langle #1 \rvert}

\newcommand{\tb}[1]{\textcolor{blue}{#1}}

\date{\today}

\begin{abstract}
Discrete dynamics arise naturally in systems with broken temporal translation symmetry and are typically described by first-order recurrence relations representing classical or quantum Markov chains.
When memory effects induced by hidden degrees of freedom are relevant, however, higher-order discrete evolution equations are generally required.
Focusing on linear dynamics, we identify a well-delineated weak-memory regime where such equations can, on an intermediate time scale, be systematically reduced to a unique first-order counterpart acting on the same state space.
We formulate our results as a mathematical theorem and work out two examples showing how they can be applied to stochastic Floquet dynamics under coarse-grained and quantum collisional models. 
\end{abstract}

\maketitle

\begin{figure}
\includegraphics[width=\linewidth]{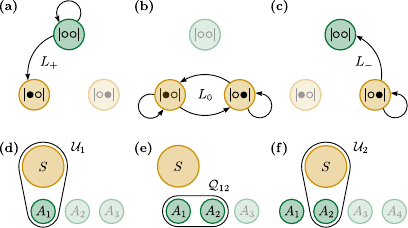}
\caption{\textbf{Top:}
Mesoscopic charge pump. 
The device consists of two quantum dots, each coupled to a thermochemical reservoir, and admits three microstates corresponding to the left dot, right dot, or neither being occupied.
Charge pumping proceeds in a three-stroke cycle: a particle is absorbed by the left dot (1a), tunnels between the dots (1b), and is ejected from the right dot (1c), with respective probabilities $L_+$, $L_0$ and $L_-$. 
An observer that measures the total charge of the system distinguishes two mesostates, whose dynamics are non-Markovian.
\textbf{Bottom:}
Collisional model with memory. 
A quantum system $S$ interacts sequentially with a stream of identical ancillas $A_i$. 
Ancilla $A_1$ first collides with the system (2a), then with $A_2$ (2b), before leaving the scattering region (2c).
System-ancilla and ancilla-ancilla interactions are described by bipartite maps $\mU_i$ and $\mQ_{ij}$, respectively.
\label{fig:Int}
}

\end{figure}

The dynamical principles of classical and quantum mechanics imply that physical systems across all scales evolve continuously in time. 
Nonetheless, processes with broken temporal translation symmetry can often be naturally understood as sequences of discrete steps. 
Isolated systems subject to rapidly oscillating fields, for example, admit a stroboscopic description in terms of time-independent effective Floquet Hamiltonians that can be engineered through the applied driving, making it possible to realize phenomena with no equilibrium counterparts~\cite{bukovUniversalHighfrequencyBehavior2015,
eckardtColloquiumAtomicQuantum2017,
okaFloquetEngineeringQuantum2019}.  
In open systems, periodic driving typically leads to the formation of limit cycles, where state variables converge to stable fixed points in stroboscopic time~\cite{moskaletsFloquetScatteringTheory2002,
rayStochasticThermodynamicsPeriodically2017,
menczelLimitCyclesPeriodically2019,
moriFloquetStatesOpen2023}. 
This behavior arises from dissipation balancing energy absorption and can be observed, for instance, in mesoscopic devices such as stochastic pumps, which generate directed currents through cyclic variations of control parameters~\cite{sinitsynUniversalGeometricTheory2007,
rahavDirectedFlowNonadiabatic2008,
espositoStochasticThermodynamicsHidden2015,
razMimickingNonequilibriumSteady2016,
funoShortcutsAdiabaticPumping2020}, see Fig.~\ref{fig:Int}.

Besides arising naturally from periodic modulations of continuous dynamics, discrete evolution equations also serve as powerful tools to disentangle the complexity of systems with large numbers of interacting degrees of freedom.
Discrete-time lattice models, such as cellular automata and quantum circuits, for instance, offer elegant frameworks to explore how macroscopic features of many-body systems such as entanglement spreading and hydrodynamic behavior emerge from local interactions~\cite{nahumQuantumEntanglementGrowth2017,
nahumOperatorSpreadingRandom2018,
vonkeyserlingkOperatorHydrodynamicsOTOCs2018,
khemaniOperatorSpreadingEmergence2018,
chanSolutionMinimalModel2018,
rakovszkyDiffusiveHydrodynamicsOutofTimeOrdered2018,
rakovszkySubballisticGrowthRenyi2019,
bertiniEntanglementSpreadingMinimal2019,
klobasExactThermalizationDynamics2021,
klobasEntanglementDynamicsRule2021,
klobasExactRelaxationGibbs2021,
bucaRule54ExactlySolvableModel2021,
bertiniExactQuenchDynamics2024,
kimCircuitsSimplePlatform2025}.
In a similar spirit, collisional models enable an elegant description of open quantum systems, even beyond the weak-coupling regime, by replacing continuous environments with discrete collections of ancillas that interact sequentially with the observable 
system~\cite{karevskiQuantumNonequilibriumSteady2009,
strasbergQuantumInformationThermodynamics2017,
strasbergRepeatedInteractionsQuantum2019,
seahCollisionalQuantumThermometry2019,
rodriguesThermodynamicsWeaklyCoherent2019}, see Fig.~\ref{fig:Int}.

A prototypical discrete evolution equation is
\begin{equation}\label{eq:adiabatic_dynamics}
    X_{n+1} = \sV X_{n},
\end{equation}
where the vector $X_n\in\mathbb{C}^N$ characterizes the state of the system after $n$ time steps, for example, in terms of a probability distribution, and $\sV\in\mathbb{C}^{N\times N}$ is the one-step propagator or generator. 
Such first-order recurrence relations provide suitable models for systems whose future is, at any point in time, fully determined by their present state. 
This condition is satisfied if all relevant degrees of freedom are accessible to the observer.   
In particular in mesoscopic systems, however, it is typically only possible to resolve coarse grained  variables, while details of internal and environmental structures remain inaccessible~\cite{pigolottiCoarseGrainingMaster2008,
espositoStochasticThermodynamicsCoarse2012,
yvinecAdiabaticReductionModel2014,
hummerOptimalDimensionalityReduction2015,
wachtelThermodynamicallyConsistentCoarse2018,
lapollaManifestationsProjectionInducedMemory2019,
strasbergNonMarkovianityNegativeEntropy2019,
busielloCoarsegrainedEntropyProduction2020,
seiferthCoarseGrainingBiochemical2020,
loosIrreversibilityHeatAndInformation2020,
straubeRapidOnset2020, 
ayazNonMarkovianModelingProtein2021,
hartichEmergentMemoryKinetic2021,
lapollaToolboxQuantifyingMemory2021,
hartichViolationLocalDetailed2023,
coghiCurrentFluctuations2024,
blomMilestoningEstimatorsDissipation2024,
vollmarModelfreeInferenceMemory2024,
zhaoEmergenceMemoryEquilibrium2024,
wisniewskiMemoryinducedCurrentReversal2025a}.
These hidden degrees of freedom can, if they evolve sufficiently slowly, retain information about the observable system over several time steps, while constantly influencing its dynamics.
Accounting for this effect requires more general evolution equations of the form 
\begin{equation}
    \label{eq:starting_point}
    X_{n+1} = \sV X_n + \sum\nolimits_{m=1}^{n} \sK_m X_{n-m}, 
\end{equation}
where the memory kernel $\sK_n$ can either be extracted from experimental or simulation data, or derived by projecting the state space of a microscopic model onto a lower-dimensional space of mesostates~\cite{lesnickiMolecularHydrodynamics2016, jungIterativeReconstruction2017, meyerNumericalProcedure2020, vrugtProjectionOperatorsStatistical2020,
klippensteinIntroducingMemoryCoarseGrained2021,
schillingCoarsegrainedModellingOut2022,
hilderQuantitativeCoarseGrainingMarkov2024}. 
We assume throughout that neither the one-step propagator nor the memory the kernel depend explicitly on time. 
This condition is satisfied whenever the underlying micro-dynamics are stroboscopically homogeneous, as is the case for stationary and periodically driven processes.

Higher-order recurrence relations of the type of Eq.~\eqref{eq:starting_point} can provide significantly more accurate models for systems with hidden degrees of freedom than their first-order counterparts, especially when high temporal resolution is required and memory effects cannot be suppressed by controlling the time step.
However, they are also less physically transparent and more difficult to solve both analytically and numerically, especially when the accessible state space remains large. 
Here, we show how this trade-off can be systematically overcome in settings where memory effects play a subdominant role. 
To this end, we reconstruct a recently developed theory of continuous weak-memory in discrete time, thereby substantially broadening the scope of this framework~\cite{
brandnerDynamicsMicroscaleNanoscale2025a,
brandnerDynamicsMicroscaleNanoscale2025}.

The key idea of our approach is to construct an effective generator $\sG\in\mathbb{C}^{N\times N}$ such that the solution $Y_n\in\mathbb{C}^N$ of the first-order recurrence relation   
\begin{equation}\label{eq:LTApprox}
	Y_{n+1} = \sG Y_n
\end{equation}
converges to the solution of Eq.~\eqref{eq:starting_point} at sufficiently long times. 
The short-time dynamics of the system are described as an effectively instantaneous change of its initial state, $Y_0 = \sD X_0$, induced by the slippage matrix $\sD \in \mathbb{C}^{N \times N}$, which, for any $X_0$, has to satisfy 
\begin{equation}\label{eq:SMAsymp}
	\lim_{n\rightarrow\infty} \sG^{-n} X_n = \sD X_0.
\end{equation}
Here, $\sG$ and $\sD$ must be non-singular to ensure that the long-time approximation $Y_n$ is non-trivial for any $X_0\neq 0$.
The physical intuition behind this ansatz is that, after a brief transient period, all hidden degrees of freedom reach a near-stationary state, which then evolves almost adiabatically under the influence of the observable ones.  
As a result, the backflow of information from inaccessible to accessible components of the system gradually diminishes and the impact of memory effects becomes insignificant.  

The validity of this picture depends on three characteristic time scales, which we define by imposing the bounds 
\begin{equation}\label{eq:WMC1}
	 \nrm{\sV^{-1}}\leq 1/v, \qquad \nrm{\sK_n}\leq Mk^{n-1}
\end{equation}
on the free generator and the memory kernel, where $\nrm{\cdot}$ denotes the spectral norm. 
Here, $v$ corresponds to the slowest mode of the adiabatic dynamics, which would be observed if the evolution of all hidden degrees of freedom was frozen. 
The parameters $M$ and $k < 1$ quantify, respectively, the coupling strength between accessible and inaccessible parts of the system and the relaxation rate of the latter.
Thus, when $M$ and $k$ are small compared to $v$, the hidden degrees of freedom evolve nearly autonomously, with the observable ones acting only as a weak and slow perturbation. 
Under these conditions, memory effects are expected to be subdominant and an effective generator satisfying Eq.~\eqref{eq:SMAsymp} may exist. 

Our main achievement is to translate the qualitative analysis above into a mathematical theorem, which can be formulated as follows.  
If the parameters $v, M$ and $k$ entering the bounds \eqref{eq:WMC1} satisfy
\begin{equation}\label{eq:WMC2}
	k<v, \quad
	M < M^\ast = 
		\begin{cases}
			(v-k)^2/4, & v+k <2\\
			(1-k)(v-1),& v+k \geq 2 
		\end{cases},
\end{equation}
there exists a unique effective generator $\sG$ such that, for any initial state $X_0$, the long-time approximation $Y_n$ of the solution $X_n$ of Eq.~\eqref{eq:starting_point} satisfies
\begin{equation}\label{eq:BndErrLTA}
	\abs{X_n - Y_n}\leq \frac{\zeta-k}{\eta-\zeta}\abs{X_0}\zeta^n
\end{equation}
with
\begin{align}
	\zeta & = (v+k)/2 - \sqrt{(v-k)^2/4 -M} < 1,\\
	\eta & = (v+k)/2 + \sqrt{(v-k)^2/4- M}
\end{align}
and $\abs{\cdot}$ indicating the Euclidean norm. 
The corresponding slippage matrix is given by the formula
\begin{equation}\label{eq:SMDef}
	\sD^{-1} = 1 + \sum\nolimits_{m=1}^\infty\sum\nolimits_{n=1}^\infty \sH^{-m}\sK_{m+n-1}\sG^{-n},
\end{equation}
where $\sG$ and $\sH$ are the unique solutions of the fixed-point equations
\begin{align}
	\label{eq:GFPE}
	\sG & = \sV + \sum\nolimits_{n=1}^\infty \sK_n \sG^{-n} = T(\sG),\\
	\label{eq:HFPE}
	\sH & = \sV + \sum\nolimits_{n=1}^\infty \sH^{-n}\sK_n  = U(\sH)
\end{align}
in the set $	B=\bigl\{\sX\in\mathbb{C}^{N\times N} \; : \; \nrm{\sX^{-1}}\leq 1/\eta\bigr\}$.

The proof of this theorem and the required mathematical methods are discussed in the Supplemental Material \cite{meyerWeakMemoryDynamicsDiscrete2025}\nocite{agarwalFixedPointTheory2018,hornMatrixAnalysis2013,rivasOpenQuantumSystems2012,
burtonVolterraIntegralDifferential2005}. Here, we only add three remarks. First, Eqs.~\eqref{eq:WMC1} and \eqref{eq:WMC2} define the weak-memory regime for discrete dynamics, where our theory is applicable without further restrictions.
These conditions do not require a strong separation of time scales, but can be met even when $M, k$ and $v$ are of the same order of magnitude.
In such situations, phenomenological methods such as the standard Born-Markov approximation are typically not reliable \cite{strunzConvolutionlessNonMarkovian2004, flemingAccuracyPerturbativeMaster2011, hartmannAccuracyAssessmentPerturbative2020, tellobreuerBenchmarkingQuantumMaster2024, sartipiCanonicallyConsistentQuantumMaster2026}.

Second, Eq.~\eqref{eq:BndErrLTA} provides a strong bound on the error of the long-time approximation, thereby enforcing the uniqueness of the effective generator. 
That is, while there may exist a whole family of generators that define a non-singular slippage matrix, there is only one such pair whose corresponding long-time approximation satisfies
\begin{equation}
	\lim_{n\rightarrow\infty} \abs{X_n - Y_n}\zeta^{-n} < \infty
\end{equation}
for any $X_0$. 
Finally, Eqs.~\eqref{eq:SMDef}, \eqref{eq:GFPE} and \eqref{eq:HFPE} provide an explicit scheme to construct this optimal approximation to arbitrary accuracy, which would not be possible with standard weak-coupling expansions whose convergence properties are not generally understood.
To this end, the fixed-point equations \eqref{eq:GFPE} and \eqref{eq:HFPE} may be solved with a suitable ansatz or by iteration, where the latter approach exploits the fact that $T$ and $U$ are contractions on $B$ under the metric $d(\sX,\sY)=\nrm{\sX-\sY}$. 

As a first application, we show how theory can be employed in the coarse graining of Markov processes. 
We focus on systems with a finite space of microstates $\mathbb{M}=\{1,\dots,M\}$ and denote by $P_n=[P_n^1,\dots,P_n^M]$ the time dependent vector that assigns an occupation probability to every microstate.
Dynamics are governed by the discrete-time master equation 
$P_{n+1} = \sL P_n$, 
where $(\sL)_{ij}\geq 0$ is the probability to transition from the microstates $i$ to the microstate $j$ in one step.    
A coarse graining is induced by a partitioning of $\mathbb{M}$ into $N$ disjoint subsets $\mathbb{M}_\alpha$, which correspond to the operationally distinguishable configurations of the system. 
The occupation probabilities $X_n^\alpha = \sum_{i\in\mathbb{M}_\alpha}P^i_n$ of these mesostates are collected in the lumped vector $X_n=[X_n^1,\dots,X_n^N]$. 
Under minor technical conditions, this vector follows an evolution equation of the form \eqref{eq:starting_point}, where $\sV$ and $\sK_n$ can be expressed in terms of the microscopic transition matrix $\sL$ using projection operator techniques~\cite{vrugtProjectionOperatorsStatistical2020,
klippensteinIntroducingMemoryCoarseGrained2021,
schillingCoarsegrainedModellingOut2022,
hilderQuantitativeCoarseGrainingMarkov2024}. 

Our master theorem extends this framework by establishing rigorous conditions for the existence of, and systematic procedures to construct, an effective generator of the lumped dynamics.
These results apply to broad classes of physical systems, including in particular ones with continuous underlying dynamics.
Systems such as electronic nano-devices, for example, are typically  described by a continuous-time master equation~\cite{seifertStochasticThermodynamicsFluctuation2012,
benentiFundamentalAspectsSteadystate2017},
$\dot{p}_t = \sW_t p_t.$
Here, $p_t = [p^1_t,\dots, p^M_t]^\trans$ is a vector of occupation probabilities, $(\sW_t)_{ij} \geq 0$ is the transition rate from the micro\-state $j$ to the micro\-state $i$, which becomes time dependent under external driving, and $(\sW_t)_{ii} = -\sum_{j\neq i} (\sW_t)_{ji}$ to ensure probability conservation. 
If the driving is periodic, $\sW_{t+\tau} = \sW_t$, Floquet's theorem implies that the stroboscopic vector $P_n = p_{n\tau}$ follows a discrete-time master equation with the transition matrix $\sL = \exp_{\mathrm{O}}[\int_0^\tau dt\;  \sW_t]$.
This observation shows that our theory opens a path to systematically derive effective generators for coarse-grained, periodically modulated dynamics, a task that could not be addressed with the existing weak-memory theory for continuous dynamics, and, more generally, poses a significant challenge without additional assumptions on the driving time scale or the use of uncontrolled approximations \cite{mozgunovCompletelyPositiveMaster2020,
nathanUniversalLindbladEquation2020,
schnellThereFloquetLindbladian2020,
hotzCoarsegrainingMasterEquation2021, 
moriStrongMarkovDissipation2025}.

Two remarks are in order before we move on to a simple example.
First, if $\sL$ is a stochastic matrix, as assumed above, then so is the free generator $\sV$ of the lumped dynamics. 
Consequently, the Perron-Frobenius theorem implies $v\leq 1$ for this entire class of models. 
Second, while we have so far focused on classical Markov jump processes, our conclusions extend directly to open quantum systems with a finite-dimensional Hilbert space, upon replacing probability vectors, transition matrices, and rate matrices by density matrices, quantum channels, and Lindblad generators, respectively.

\begin{figure}
    \includegraphics[width=\linewidth]{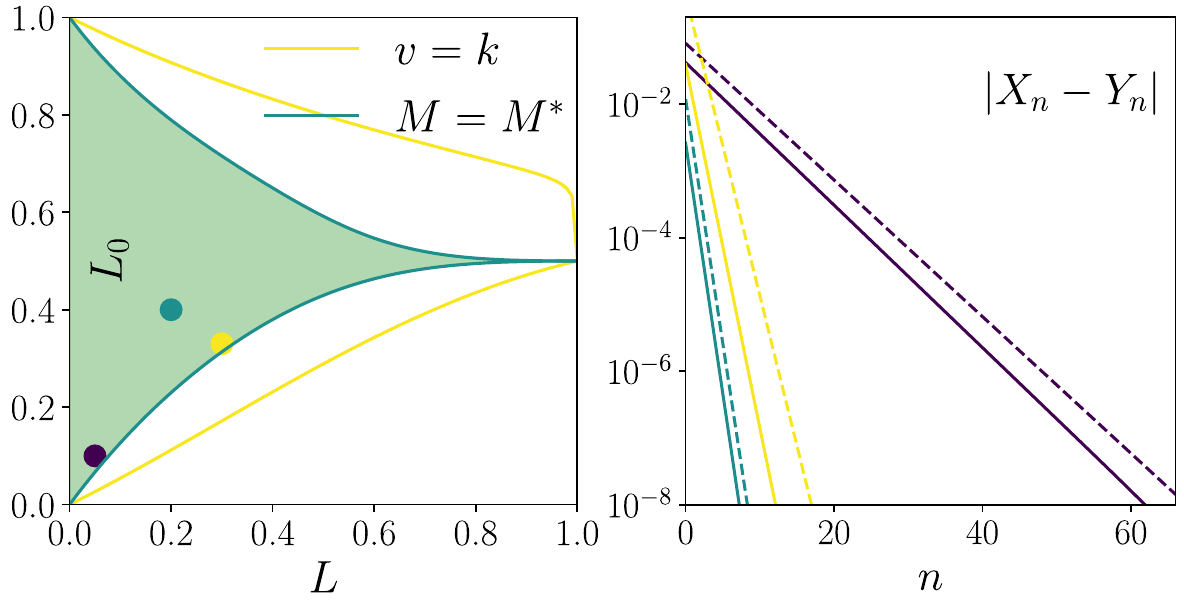}
    \caption{Mesoscopic charge pump. 
    \textbf{Left:}
    For $L_+ = L_- = L$, the shaded area indicates the region in the parameter space of the model,
    where the weak-memory conditions \eqref{eq:WMC1} and \eqref{eq:WMC2} are satisfied with
    $v=1/\nrm{\sV^{-1}}$ and $M=\nrm{\sK_1}$. 
    \textbf{Right:}
    Error of the long-time approximation $Y_n=\sG^n\sD X_0$ with respect to the exact solution
    $X_n$ of Eq.~\eqref{eq:starting_point} (solid), for $X_0 = [1,0]^\trans$ and selected
    parameter sets, indicated with dots in the left panel.
    For comparison, we also show the error bound \eqref{eq:BndErrLTA} (dashed). 
    }
    \label{fig:Cpm}
\end{figure}

We now focus on the charge-pump model of Fig.~\ref{fig:Int}.
A complete operation cycle consists of the three strokes shown in panels (a), (b), (c), which are described by the stochastic matrices $\sL_+$, $\sL_0$, $\sL_-$, respectively. 
The microscopic transition matrix is given by $\sL=\sL_-\sL_0\sL_+$. 
For an observer monitoring only the total charge of the two quantum dots, the system is described by the lumped probability vector $X_n = [X^\circ_n, X^\bullet_n]^\trans$, where $X^\circ_n = P^{\circ\circ}_n$ and $X^\bullet_n = P^{\circ\bullet}_n + P^{\bullet\circ}_n$. 
If the initial charge distribution is symmetric, as we assume here only for the sake of simplicity, $P^{\circ\bullet}_0 = P^{\bullet\circ}_0$, this vector follows Eq.~\eqref{eq:starting_point} with 
\begin{align}
	\sV & = 1 + L_+(1-L_0L_-)\sJ_1 + \frac{L_-}{2}\sJ_2,\\
	\sK_n & = \frac{L_+L_-(1-2L_0+2L_-)k^n}{2-L_-}\sJ_1 
		- \frac{L_-^ 2k^n}{2(2-L_-)}\sJ_2,
\end{align}
where $k= (1-2L_0)(2-L_-)/2$,
$\sJ_1 = \big[\begin{smallmatrix} -1 & 0\\ \hphantom{-}1 & 0 \end{smallmatrix}\big]$, 
$\sJ_2 = \big[\begin{smallmatrix} 0 & \hphantom{-}1\\ 0 & -1 \end{smallmatrix}\big]$,
and $L_+$, $L_0$ and $L_-$ are the absorption, tunneling and ejection probabilities as defined in Fig.~\ref{fig:Int}.  
For $L_0 = 1/2$, the charge distribution fully equilibrates in one cycle and the memory kernel vanishes. 
Away from this limit, the parameters $L_\pm$, which are determined by the coupling strengths between quantum dots and reservoirs, must remain sufficiently small for the weak-memory conditions \eqref{eq:WMC1} and \eqref{eq:WMC2} to hold, see Fig.~\ref{fig:Cpm}. 
In this regime, the fixed-point equation \eqref{eq:GFPE} for the effective generator can be solved using the ansatz $\sG = 1 + g_1\sJ_1 + g_2\sJ_2$, and the adjoint generator $\sH$ follows analogously. 
The long-time approximation $Y_n = \sG^n\sD X_0$ is then obtained using the formula \eqref{eq:SMDef} for the slippage matrix $\sD$~\cite{meyerWeakMemoryDynamicsDiscrete2025}. 
As Fig.~\ref{fig:Cpm} shows, the error of this approximation decays exponentially with $n$, closely following the upper bound \eqref{eq:BndErrLTA}.

\begin{figure}
    \includegraphics[width=\linewidth]{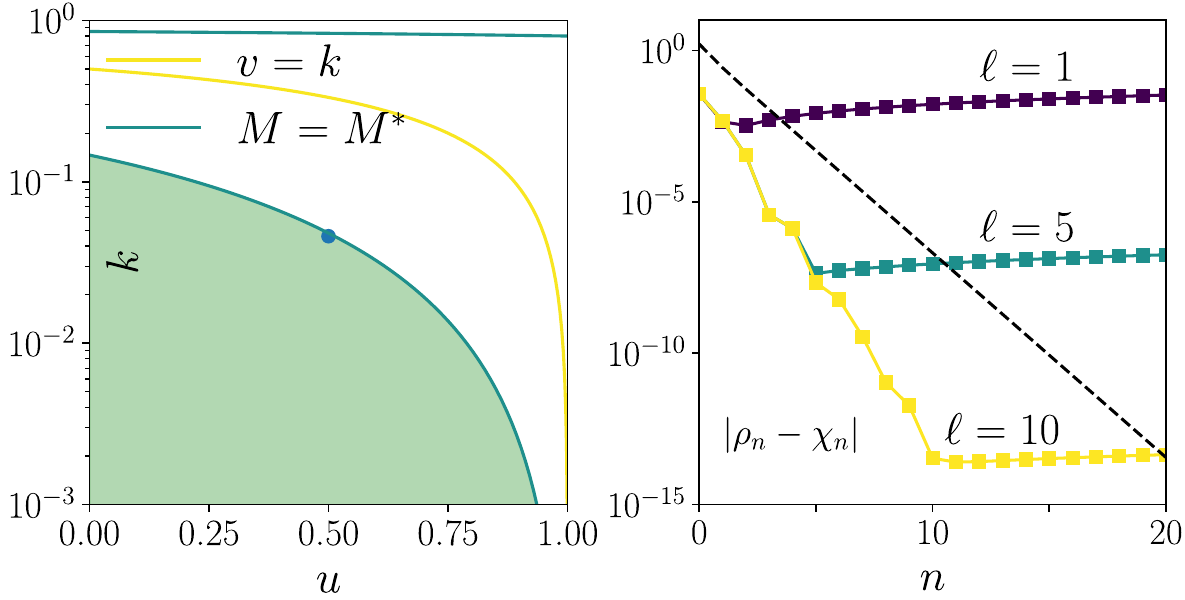}
    \caption{Collisional model. 
    \textbf{Left:}
    Weak-memory regime in the parameter space defined by the swap probabilities $u$ and $k$, 
    with $v = (1-u)(1-k)$ and $M=k(1-k)$. 
    The ancillas are initially in fully mixed states. 
    \textbf{Right:}
    Error of the long-time approximation \eqref{eq:LTACM} for different approximate generators 
    $\mG_\ell = T(\mG_{\ell-1})$ (solid), together with the bound 
    \eqref{eq:BndLTACM} (dashed). 
    For all plots, we used the model parameters indicated by the dot in the left panel, and 
    set $\rho_0 = \ket{0_S}\bra{0_S}$.
    \label{fig:Clm}
    }
\end{figure}

As our second main application, we consider collisional models, which offer a versatile platform to describe open or monitored quantum dynamics in discrete time~\cite{ciccarelloQuantumCollisionModels2022}. 
As illustrated in Fig.~1, such models consist of a system of interest that couples sequentially to a collection of identical environmental units, or ancillas.
Memory effects can be added to this framework, for instance, through interactions between successive ancillas. 
The state of the system after $n$ collisions is then given by 
\begin{equation}
	\rho_n = \tr_A[\mU_n \;\cdots\; \mU_2\mQ_{12}\mU_1(\rho_0 \otimes\xi^n)],
\end{equation}
where $\xi^n = \bigotimes_{i=1}^n\xi_i$ is the initial state of the ancillas, and $\tr_A[\circ]$ indicates the partial trace over their joint Hilbert space. 
The maps $\mU_i$ and $\mQ_{ij}$ describe bipartite interactions between the $i^\mathrm{th}$ ancilla and the system, and between the $i^\mathrm{th}$ and the $j^\mathrm{th}$ ancillas, respectively. 

For concreteness, we now assume that the system and the ancillas are qubits with ground and excited states $\ket{0_i}$ and $\ket{1_i}$, where $i=S,1,\dots,n$. 
System-ancilla and ancilla-ancilla interactions are described as unitary and incoherent partial swaps, 
\begin{equation}
	\mU_i^{\vphantom{\dagger}}\circ = \sU_i^{\vphantom{\dagger}}\circ\sU_i^\dagger, \quad
	\mQ_{ij}^{\vphantom{\dagger}}\circ= (1-k)\circ 
		+ k\sS_{ij}^{\vphantom{\dagger}}\circ\sS_{ij}^{\vphantom{\dagger}},
\end{equation}
where $\sU_i = \sqrt{1-u} - \mathrm{i} \sqrt{u}\sS_{Si}$ and $0<u,k<1$. 
The swap matrices are defined such that $\sS_{ij}\ket{\alpha_i,\beta_j} = \ket{\beta_i,\alpha_j}$ for $\alpha,\beta\in\{0,1\}$. 
If all ancillas are initially prepared in the same state, the dynamics of this model are governed by the recursion relation
\begin{equation}\label{eq:EOMCM}
	\rho_{n+1} = \mK_0 \rho_n + \sum\nolimits_{m=1}^n \mK_{m}\rho_{n-m} + \mF_{n+1}\rho_0,
\end{equation}
with $\mK_n = (1/k-1)\mF_{n+1}$ and $\mF_n\circ = k^n\tr_1^{\vphantom{n}}[\mU^n_1(\circ\otimes\xi_1^{\vphantom{n}})]$, see Refs.~\cite{meyerWeakMemoryDynamicsDiscrete2025,ciccarelloQuantumCollisionModels2022}. 
Since $\mF_n$ is a linear map on the operator space of the system, it can be represented as a four-dimensional matrix \footnote{In our discussion of collisional models, we formally treat density matrices as vectors and quantum maps as matrices. 
For this approach to be consistent, $\abs{\cdot}$ must, when applied to a density matrix, be evaluated as the Hilbert-Schmidt norm; $\nrm{\cdot}$ denotes the operator norm induced by $\abs{\cdot}$ throughout.}.
Thus, Eq.~\eqref{eq:EOMCM} has the structure of Eq.~\eqref{eq:starting_point}, up to an inhomogeneous term, which can, however, be accounted for.  
If $\mG$ and $\mD$ are the effective generator and slippage map of the homogeneous dynamics, then the long-time approximation of $\rho_n$ reads~\cite{meyerWeakMemoryDynamicsDiscrete2025}
\begin{equation}\label{eq:LTACM}
	\chi_n = \mG^n\mD\rho_0 + \sum\nolimits_{m=1}^n\mG^{n-m}\mD\mF_m\rho_0. 
\end{equation}
To evaluate this expression, we solve the fixed-point equations \eqref{eq:GFPE} and \eqref{eq:HFPE} numerically by iteration using the initial guesses $\mG_0 = \mH_0 = \mK_0$; the corresponding slippage map then follows from Eq.~\eqref{eq:SMDef}. 
As we show in Fig.~\ref{fig:Clm}, the error of the resulting approximations first decays exponentially with time, obeying the bound 
\begin{equation}\label{eq:BndLTACM}
	\abs{\rho_n-\chi_n}\leq \frac{\zeta-k}{\eta-\zeta}\abs{\rho_0} 
		\zeta^n\Bigl(1 + \sum\nolimits_{m=1}^n \nrm{\mF_m}\zeta^{-m}\Bigr),
\end{equation}
which follows from Eq.~\eqref{eq:BndErrLTA} and holds for the exact effective generator and slippage map~\cite{meyerWeakMemoryDynamicsDiscrete2025}, before approaching a plateau whose value decreases exponentially with the number of iterations used to determine the generator. 

This observation shows that iterating the fixed-point equations \eqref{eq:GFPE} and \eqref{eq:HFPE} provides an efficient numerical method for determining the effective generator and the associated slippage matrix.
However, due to their nonlinear structure, these equations generally do not provide a convenient starting point for analytical perturbation schemes.
For this purpose, it is more instructive to focus instead on the $\sE_n\in\mathbb{C}^{N\times N}$, which, for any given $\sG$, is uniquely defined by requiring that the inhomogeneous recurrence relation
\begin{equation}
	X_{n+1} = \sG X_n + \sE_n X_0
\end{equation}
is equivalent to Eq.~\eqref{eq:starting_point}. 
If the weak-memory conditions \eqref{eq:WMC1} and \eqref{eq:WMC2} are fulfilled, and $\sG$ is the optimal effective generator of our master theorem, this function satisfies the linear summation equation
\begin{equation}
	\sE_n = -\sum\nolimits_{m=n+1}^\infty\Bigl(
		\sK_{m} + \sum\nolimits_{l=1}^{m-1}\sE_{l-1}\sK_{m-l}\Bigr)\sV^{n-m}, 
\end{equation}
which can be solved by iteration. 
In this way, the optimal memory function $\sE_n$ of a given model can be systematically constructed from an initial guess $\sE^0_n$. 
This initial guess may either be set to zero, or derived from an exact solution of Eq.~\eqref{eq:starting_point} for a suitable limiting case of the model.
If the free solution satisfies the bound $\nrm{\sE^0_n}\leq S\sigma^n$ with $\zeta\leq\sigma<\min\{\eta,1\}$ and $S\geq \sigma -k$, the iteration converges. 
Once the optimal memory function has been determined, the corresponding effective generator and slippage matrix follow from the relations
\begin{equation}
\sG = \sV - \sE_0, \qquad
\sD = 1 + \sum\nolimits_{n=1}^\infty \sG^{-n}\sE_{n-1}.
\end{equation}
For the trivial initial guess $\sE^0_n = 0$, for instance, we obtain an expansion of $\sG$ in powers of the coupling strength $M$.

These results, whose derivation we provide in Ref.~\cite{meyerWeakMemoryDynamicsDiscrete2025}, complete our theory of weak memory effects in discrete time.
Before closing, we stress that, while the above framework achieves a high degree of structural uniformity with the one recently developed for continuous dynamics, our results could not be derived from that theory, since the exact discretization of evolution equations with memory is a non-trivial procedure. In particular, there is no simple correspondence between the discrete one-step propagator and memory kernel and their continuous counterparts. 
More generally, our work provides a universal basis to explore the potentially rich physics of dynamical memory in settings with inherently discrete dynamics, such as many-body circuit and collisional models. 
On the applied side, it offers new tools for the extrapolation of short-time data obtained from experiments or molecular dynamics simulations \cite{jungIterativeReconstructionMemory2017,
klippensteinIntroducingMemoryCoarseGrained2021a,
vroylandtLikelihoodbasedNonMarkovianModels2022,
tepperAccurateMemoryKernel2024,
daltonMemoryFrictionNanoscale2025,
satoreMarkovTypeStateModels2025}.
The applicability of our framework is determined solely by the weak-memory conditions \eqref{eq:WMC1} and \eqref{eq:WMC2}. 
It thus extends beyond the standard weak-coupling and quasi-Markovian limits, which impose stronger and less rigorous constraints on system parameters.
At the same time, our results provide a well-defined starting point for future research seeking to systematically characterize strong-memory dynamics.

\emph{Acknowledgments.--}
This work was supported by the Medical Research Council (Grants No. MR/S034714/1 and MR/Y003845/1) and the Engineering and Physical Sciences Research Council (Grant No. EP/V031201/1).

\emph{Data availability.--} All data that support the findings of this study are included within the article and the Supplemental Material. 

\let\oldaddcontentsline\addcontentsline
\renewcommand{\addcontentsline}[3]{}
\bibliography{manuscript.bib}

@article{strunzConvolutionlessNonMarkovian2004,
	title={Convolutionless non-Markovian master equations and quantum trajectories: Brownian motion},
	author={Strunz, Walter T and Yu, Ting},
	journal={Phys. Rev. A},
	volume={69},
	number={5},
	pages={052115},
	year={2004},
	publisher={APS}
}

@book{agarwalFixedPointTheory2018,
  title = {Fixed {{Point Theory}} in {{Metric Spaces}}},
  author = {Agarwal, P. and Jleli, M. and Samet, B.},
  year = 2018,
  edition = {1st},
  publisher = {Springer Nature},
  address = {Singapore},
  isbn = {978-981-13-2912-8}
}

@article{ayazNonMarkovianModelingProtein2021,
  title = {Non-{{Markovian}} Modeling of Protein Folding},
  author = {Ayaz, Cihan and Tepper, Lucas and Br{\"u}nig, Florian N. and Kappler, Julian and Daldrop, Jan O. and Netz, Roland R.},
  year = 2021,
  journal = {Proc. Natl. Acad. Sci. U.S.A.},
  volume = {118},
  pages = {e2023856118},
  doi = {10.1073/pnas.2023856118}
}

@article{benentiFundamentalAspectsSteadystate2017,
  title = {Fundamental Aspects of Steady-State Conversion of Heat to Work at the Nanoscale},
  author = {Benenti, Giuliano and Casati, Giulio and Saito, Keiji and Whitney, Robert~S.},
  year = 2017,
  journal = {Physics Reports},
  volume = {694},
  pages = {1--124},
  doi = {10.1016/j.physrep.2017.05.008}
}

@article{bertiniEntanglementSpreadingMinimal2019,
  title = {Entanglement {{Spreading}} in a {{Minimal Model}} of {{Maximal Many-Body Quantum Chaos}}},
  author = {Bertini, Bruno and Kos, Pavel and Prosen, Toma{\v z}},
  year = 2019,
  journal = {Phys. Rev. X},
  volume = {9},
  pages = {021033},
  doi = {10.1103/PhysRevX.9.021033}
}

@article{bertiniExactQuenchDynamics2024,
  title = {Exact {{Quench Dynamics}} of the {{Floquet Quantum East Model}} at the {{Deterministic Point}}},
  author = {Bertini, Bruno and De Fazio, Cecilia and Garrahan, Juan P. and Klobas, Katja},
  year = 2024,
  journal = {Phys. Rev. Lett.},
  volume = {132},
  pages = {120402},
  doi = {10.1103/PhysRevLett.132.120402}
}

@article{blomMilestoningEstimatorsDissipation2024,
  title = {Milestoning Estimators of Dissipation in Systems Observed at a Coarse Resolution},
  author = {Blom, Kristian and Song, Kevin and Vouga, Etienne and Godec, Alja{\v z} and Makarov, Dmitrii E.},
  year = 2024,
  journal = {Proc. Natl. Acad. Sci. U.S.A.},
  volume = {121},
  pages = {e2318333121},
  doi = {10.1073/pnas.2318333121}
}

@article{brandnerDynamicsMicroscaleNanoscale2025,
  title = {Dynamics of Microscale and Nanoscale Systems in the Weak-Memory Regime: {{A}} Mathematical Framework beyond the {{Markov}} Approximation},
  author = {Brandner, Kay},
  year = 2025,
  journal = {Phys. Rev. E},
  volume = {111},
  pages = {014137},
  doi = {10.1103/PhysRevE.111.014137}
}

@article{brandnerDynamicsMicroscaleNanoscale2025a,
  title = {Dynamics of {{Microscale}} and {{Nanoscale Systems}} in the {{Weak-Memory Regime}}},
  author = {Brandner, Kay},
  year = 2025,
  journal = {Phys. Rev. Lett.},
  volume = {134},
  pages = {037101},
  doi = {10.1103/PhysRevLett.134.037101}
}

@article{bukovUniversalHighfrequencyBehavior2015,
  title = {Universal High-Frequency Behavior of Periodically Driven Systems: From Dynamical Stabilization to {{Floquet}} Engineering},
  author = {Bukov, Marin and D'Alessio, Luca and Polkovnikov, Anatoli},
  year = 2015,
  journal = {Adv. Phys.},
  volume = {64},
  pages = {139--226},
  doi = {10.1080/00018732.2015.1055918}
}

@book{burtonVolterraIntegralDifferential2005,
  title = {Volterra {{Integral}} and {{Differential Equations}}},
  author = {Burton, T.},
  year = 2005,
  series = {Mathematics in {{Science}} and {{Engineering}}},
  edition = {2nd},
  volume = {202},
  publisher = {Elsevier},
  address = {Amsterdam},
  isbn = {978-0-444-51786-9}
}

@article{busielloCoarsegrainedEntropyProduction2020,
  title = {Coarse-Grained Entropy Production with Multiple Reservoirs: {{Unraveling}} the Role of Time Scales and Detailed Balance in Biology-Inspired Systems},
  shorttitle = {Coarse-Grained Entropy Production with Multiple Reservoirs},
  author = {Busiello, Daniel M. and Gupta, Deepak and Maritan, Amos},
  year = 2020,
  journal = {Phys. Rev. Res.},
  volume = {2},
  pages = {043257},
  doi = {10.1103/PhysRevResearch.2.043257}
}

@article{chanSolutionMinimalModel2018,
  title = {Solution of a {{Minimal Model}} for {{Many-Body Quantum Chaos}}},
  author = {Chan, Amos and De Luca, Andrea and Chalker, J. T.},
  year = 2018,
  journal = {Phys. Rev. X},
  volume = {8},
  pages = {041019},
  doi = {10.1103/PhysRevX.8.041019}
}

@article{ciccarelloQuantumCollisionModels2022,
  title = {Quantum Collision Models: {{Open}} System Dynamics from Repeated Interactions},
  author = {Ciccarello, Francesco and Lorenzo, Salvatore and Giovannetti, Vittorio and Palma, G. Massimo},
  year = 2022,
  journal = {Phys. Rep.},
  volume = {954},
  pages = {1--70},
  doi = {10.1016/j.physrep.2022.01.001}
}

@article{daltonMemoryFrictionNanoscale2025,
  title = {Memory and {{Friction}}: {{From}} the {{Nanoscale}} to the {{Macroscale}}},
  author = {Dalton, Benjamin A. and Klimek, Anton and Kiefer, Henrik and Br{\"u}nig, Florian N. and Colinet, H{\'e}l{\`e}ne and Tepper, Lucas and Abbasi, Amir and Netz, Roland R.},
  year = 2025,
  journal = {Annu. Rev. Phys. Chem.},
  volume = {76},
  pages = {431--454},
  doi = {10.1146/annurev-physchem-082423-031037}
}

@article{eckardtColloquiumAtomicQuantum2017,
  title = {Colloquium: {{Atomic}} Quantum Gases in Periodically Driven Optical Lattices},
  author = {Eckardt, Andr{\'e}},
  year = 2017,
  journal = {Rev. Mod. Phys.},
  volume = {89},
  pages = {011004},
  doi = {10.1103/RevModPhys.89.011004}
}

@article{espositoStochasticThermodynamicsCoarse2012,
  title = {Stochastic Thermodynamics under Coarse Graining},
  author = {Esposito, Massimiliano},
  year = 2012,
  journal = {Phys. Rev. E},
  volume = {85},
  pages = {041125},
  doi = {10.1103/PhysRevE.85.041125}
}

@article{espositoStochasticThermodynamicsHidden2015,
  title = {Stochastic Thermodynamics of Hidden Pumps},
  author = {Esposito, Massimiliano and Parrondo, Juan M. R.},
  year = 2015,
  journal = {Phys. Rev. E},
  volume = {91},
  pages = {052114},
  doi = {10.1103/PhysRevE.91.052114}
}

@article{flemingAccuracyPerturbativeMaster2011,
  title = {Accuracy of Perturbative Master Equations},
  author = {Fleming, C. H. and Cummings, N. I.},
  year = 2011,
  journal = {Phys. Rev. E},
  volume = {83},
  doi = {10.1103/physreve.83.031117}
}

@article{funoShortcutsAdiabaticPumping2020,
  title = {Shortcuts to {{Adiabatic Pumping}} in {{Classical Stochastic Systems}}},
  author = {Funo, Ken and Lambert, Neill and Nori, Franco and Flindt, Christian},
  year = 2020,
  journal = {Phys. Rev. Lett.},
  volume = {124},
  pages = {150603},
  doi = {10.1103/PhysRevLett.124.150603}
}

@article{hartichEmergentMemoryKinetic2021,
  title = {Emergent {{Memory}} and {{Kinetic Hysteresis}} in {{Strongly Driven Networks}}},
  author = {Hartich, David and Godec, Alja{\v z}},
  year = 2021,
  journal = {Phys. Rev. X},
  volume = {11},
  pages = {041047},
  doi = {10.1103/PhysRevX.11.041047}
}

@article{hartichViolationLocalDetailed2023,
  title = {Violation of Local Detailed Balance upon Lumping despite a Clear Timescale Separation},
  author = {Hartich, David and Godec, Alja{\v z}},
  year = 2023,
  journal = {Phys. Rev. Res.},
  volume = {5},
  number = {3},
  pages = {L032017},
  doi = {10.1103/PhysRevResearch.5.L032017},
  langid = {english}
}

@article{hartmannAccuracyAssessmentPerturbative2020,
  title = {Accuracy Assessment of Perturbative Master Equations: {{Embracing}} Nonpositivity},
  author = {Hartmann, Richard and Strunz, Walter T.},
  year = 2020,
  journal = {Phys. Rev. A},
  volume = {101},
  pages = {012103},
  doi = {10.1103/PhysRevA.101.012103}
}

@article{hilderQuantitativeCoarseGrainingMarkov2024,
  title = {Quantitative {{Coarse-Graining}} of {{Markov Chains}}},
  author = {Hilder, Bastian and Sharma, Upanshu},
  year = 2024,
  journal = {SIAM J. Math. Anal.},
  volume = {56},
  pages = {913--954},
  issn = {0036-1410, 1095-7154},
  doi = {10.1137/22M1473996}
}

@book{hornMatrixAnalysis2013,
  title = {Matrix {{Analysis}}},
  author = {Horn, R. A. and Johnson, C. R.},
  year = 2013,
  edition = {2nd},
  publisher = {Cambridge University Press},
  address = {New York, NY},
  isbn = {978-0-521-83940-2}
}

@article{hotzCoarsegrainingMasterEquation2021,
  title = {Coarse-Graining Master Equation for Periodically Driven Systems},
  author = {Hotz, Ronja and Schaller, Gernot},
  year = 2021,
  journal = {Phys. Rev. A},
  volume = {104},
  pages = {052219},
  doi = {10.1103/PhysRevA.104.052219}
}

@article{hummerOptimalDimensionalityReduction2015,
  title = {Optimal {{Dimensionality Reduction}} of {{Multistate Kinetic}} and {{Markov-State Models}}},
  author = {Hummer, Gerhard and Szabo, Attila},
  year = 2015,
  journal = {J. Phys. Chem. B},
  volume = {119},
  pages = {9029--9037},
  doi = {10.1021/jp508375q}
}

@article{jungIterativeReconstructionMemory2017,
  title = {Iterative {{Reconstruction}} of {{Memory Kernels}}},
  author = {Jung, Gerhard and Hanke, Martin and Schmid, Friederike},
  year = 2017,
  journal = {J. Chem. Theory Comput.},
  volume = {13},
  pages = {2481--2488},
  doi = {10.1021/acs.jctc.7b00274}
}

@article{karevskiQuantumNonequilibriumSteady2009,
  title = {Quantum {{Nonequilibrium Steady States Induced}} by {{Repeated Interactions}}},
  author = {Karevski, Dragi and Platini, Thierry},
  year = 2009,
  journal = {Phys. Rev. Lett.},
  volume = {102},
  pages = {207207},
  doi = {10.1103/PhysRevLett.102.207207}
}

@article{khemaniOperatorSpreadingEmergence2018,
  title = {Operator {{Spreading}} and the {{Emergence}} of {{Dissipative Hydrodynamics}} under {{Unitary Evolution}} with {{Conservation Laws}}},
  author = {Khemani, Vedika and Vishwanath, Ashvin and Huse, David A.},
  year = 2018,
  journal = {Phys. Rev. X},
  volume = {8},
  pages = {031057},
  doi = {10.1103/PhysRevX.8.031057}
}

@misc{kimCircuitsSimplePlatform2025,
  title = {Circuits as a Simple Platform for the Emergence of Hydrodynamics in Deterministic Chaotic Many-Body Systems},
  author = {Kim, Sun Woo P. and H{\"u}bner, Friedrich and Garrahan, Juan P. and Doyon, Benjamin},
  year = 2025,
  number = {2503.08788},
  eprint = {2503.08788},
  publisher = {arXiv},
  archiveprefix = {arXiv}
}

@article{klippensteinIntroducingMemoryCoarseGrained2021,
  title = {Introducing {{Memory}} in {{Coarse-Grained Molecular Simulations}}},
  author = {Klippenstein, Viktor and Tripathy, Madhusmita and Jung, Gerhard and Schmid, Friederike and Van Der Vegt, Nico F. A.},
  year = 2021,
  journal = {J. Phys. Chem. B},
  volume = {125},
  pages = {4931--4954},
  doi = {10.1021/acs.jpcb.1c01120}
}

@article{klippensteinIntroducingMemoryCoarseGrained2021a,
  title = {Introducing {{Memory}} in {{Coarse-Grained Molecular Simulations}}},
  author = {Klippenstein, Viktor and Tripathy, Madhusmita and Jung, Gerhard and Schmid, Friederike and Van Der Vegt, Nico F. A.},
  year = 2021,
  journal = {J. Phys. Chem. B},
  volume = {125},
  pages = {4931--4954},
  doi = {10.1021/acs.jpcb.1c01120}
}

@article{klobasEntanglementDynamicsRule2021,
  title = {Entanglement Dynamics in {{Rule}} 54: {{Exact}} Results and Quasiparticle Picture},
  author = {Klobas, Katja and Bertini, Bruno},
  year = 2021,
  journal = {SciPost Phys.},
  volume = {11},
  pages = {107},
  doi = {10.21468/SciPostPhys.11.6.107}
}

@article{klobasExactRelaxationGibbs2021,
  title = {Exact Relaxation to {{Gibbs}} and Non-Equilibrium Steady States in the Quantum Cellular Automaton {{Rule}} 54},
  author = {Klobas, Katja and Bertini, Bruno},
  year = 2021,
  journal = {SciPost Phys.},
  volume = {11},
  pages = {106},
  doi = {10.21468/SciPostPhys.11.6.106}
}

@article{klobasExactThermalizationDynamics2021,
  title = {Exact {{Thermalization Dynamics}} in the ``{{Rule}} 54'' {{Quantum Cellular Automaton}}},
  author = {Klobas, Katja and Bertini, Bruno and Piroli, Lorenzo},
  year = 2021,
  month = apr,
  journal = {Phys. Rev. Lett.},
  volume = {126},
  number = {16},
  pages = {160602},
  issn = {0031-9007, 1079-7114},
  doi = {10.1103/PhysRevLett.126.160602},
  urldate = {2025-10-22},
  langid = {english}
}

@article{sartipiCanonicallyConsistentQuantumMaster2026,
	title={Canonically consistent quantum master equation for proton-transfer reactions},
	author={Sartipi, Zahra and Gundermann, Richard and Anders, Janet and Saalfrank, Peter},
	journal={arXiv preprint arXiv:2603.21865},
	year={2026}
}

@article{lapollaManifestationsProjectionInducedMemory2019,
  title = {Manifestations of {{Projection-Induced Memory}}: {{General Theory}} and the {{Tilted Single File}}},
  author = {Lapolla, Alessio and Godec, Alja{\v z}},
  year = 2019,
  journal = {Front. Phys.},
  volume = {7},
  pages = {182},
  doi = {10.3389/fphy.2019.00182}
}

@article{lapollaToolboxQuantifyingMemory2021,
  title = {Toolbox for Quantifying Memory in Dynamics along Reaction Coordinates},
  author = {Lapolla, Alessio and Godec, Alja{\v z}},
  year = 2021,
  journal = {Phys. Rev. Res.},
  volume = {3},
  pages = {L022018},
  doi = {10.1103/PhysRevResearch.3.L022018}
}

@article{menczelLimitCyclesPeriodically2019,
  title = {Limit Cycles in Periodically Driven Open Quantum Systems},
  author = {Menczel, Paul and Brandner, Kay},
  year = 2019,
  journal = {J. Phys. A: Math. Theor.},
  volume = {52},
  pages = {43LT01},
  doi = {10.1088/1751-8121/ab435a}
}

@misc{meyerWeakMemoryDynamicsDiscrete2025,
  title = {Weak-{{Memory Dynamics}} in {{Discrete Time}}: {{Supplemental Material}}},
  author = {Meyer, Hugues and Brandner, Kay},
  year = 2025
}

@article{moriFloquetStatesOpen2023,
  title = {Floquet {{States}} in {{Open Quantum Systems}}},
  author = {Mori, Takashi},
  year = 2023,
  journal = {Annu. Rev. Condens. Matter Phys.},
  volume = {14},
  pages = {35--56},
  doi = {10.1146/annurev-conmatphys-040721-015537}
}

@article{moriStrongMarkovDissipation2025,
  title = {Strong {{Markov Dissipation}} in {{Driven-Dissipative Quantum Systems}}},
  author = {Mori, Takashi},
  year = 2025,
  journal = {J. Stat. Phys.},
  volume = {192},
  pages = {1},
  doi = {10.1007/s10955-024-03377-7}
}

@article{moskaletsFloquetScatteringTheory2002,
  title = {Floquet Scattering Theory of Quantum Pumps},
  author = {Moskalets, M. and B{\"u}ttiker, M.},
  year = 2002,
  journal = {Phys. Rev. B},
  volume = {66},
  pages = {205320},
  doi = {10.1103/PhysRevB.66.205320}
}

@article{mozgunovCompletelyPositiveMaster2020,
  title = {Completely Positive Master Equation for Arbitrary Driving and Small Level Spacing},
  author = {Mozgunov, Evgeny and Lidar, Daniel},
  year = 2020,
  journal = {Quantum},
  volume = {4},
  pages = {227},
  doi = {10.22331/q-2020-02-06-227},
  langid = {english}
}

@article{nahumOperatorSpreadingRandom2018,
  title = {Operator {{Spreading}} in {{Random Unitary Circuits}}},
  author = {Nahum, Adam and Vijay, Sagar and Haah, Jeongwan},
  year = 2018,
  journal = {Phys. Rev. X},
  volume = {8},
  pages = {021014},
  doi = {10.1103/PhysRevX.8.021014}
}

@article{nahumQuantumEntanglementGrowth2017,
  title = {Quantum {{Entanglement Growth}} under {{Random Unitary Dynamics}}},
  author = {Nahum, Adam and Ruhman, Jonathan and Vijay, Sagar and Haah, Jeongwan},
  year = 2017,
  journal = {Phys. Rev. X},
  volume = {7},
  pages = {031016},
  doi = {10.1103/PhysRevX.7.031016}
}

@article{nathanUniversalLindbladEquation2020,
  title = {Universal {{Lindblad}} Equation for Open Quantum Systems},
  author = {Nathan, Frederik and Rudner, Mark S.},
  year = 2020,
  journal = {Phys. Rev. B},
  volume = {102},
  pages = {115109},
  doi = {10.1103/PhysRevB.102.115109}
}

@article{okaFloquetEngineeringQuantum2019,
  title = {Floquet {{Engineering}} of {{Quantum Materials}}},
  author = {Oka, Takashi and Kitamura, Sota},
  year = 2019,
  journal = {Annu. Rev. Condens. Matter Phys.},
  volume = {10},
  pages = {387--408},
  doi = {10.1146/annurev-conmatphys-031218-013423}
}

@article{pigolottiCoarseGrainingMaster2008,
  title = {Coarse Graining of Master Equations with Fast and Slow States},
  author = {Pigolotti, Simone and Vulpiani, Angelo},
  year = 2008,
  journal = {J. Chem. Phys.},
  volume = {128},
  pages = {154114},
  doi = {10.1063/1.2907242}
}

@article{rahavDirectedFlowNonadiabatic2008,
  title = {Directed {{Flow}} in {{Nonadiabatic Stochastic Pumps}}},
  author = {Rahav, Saar and Horowitz, Jordan and Jarzynski, Christopher},
  year = 2008,
  journal = {Phys. Rev. Lett.},
  volume = {101},
  pages = {140602},
  doi = {10.1103/PhysRevLett.101.140602}
}

@article{rakovszkyDiffusiveHydrodynamicsOutofTimeOrdered2018,
  title = {Diffusive {{Hydrodynamics}} of {{Out-of-Time-Ordered Correlators}} with {{Charge Conservation}}},
  author = {Rakovszky, Tibor and Pollmann, Frank and Von Keyserlingk, C. W.},
  year = 2018,
  journal = {Phys. Rev. X},
  volume = {8},
  pages = {031058},
  doi = {10.1103/PhysRevX.8.031058}
}

@article{rakovszkySubballisticGrowthRenyi2019,
  title = {Sub-Ballistic {{Growth}} of {{R\'enyi Entropies}} Due to {{Diffusion}}},
  author = {Rakovszky, Tibor and Pollmann, Frank and Von Keyserlingk, C. W.},
  year = 2019,
  journal = {Phys. Rev. Lett.},
  volume = {122},
  pages = {250602},
  doi = {10.1103/PhysRevLett.122.250602}
}

@article{rayStochasticThermodynamicsPeriodically2017,
  title = {Stochastic Thermodynamics of Periodically Driven Systems: {{Fluctuation}} Theorem for Currents and Unification of Two Classes},
  author = {Ray, Somrita and Barato, Andre C.},
  year = 2017,
  journal = {Phys. Rev. E},
  volume = {96},
  pages = {052120},
  doi = {10.1103/PhysRevE.96.052120}
}

@article{razMimickingNonequilibriumSteady2016,
  title = {Mimicking {{Nonequilibrium Steady States}} with {{Time-Periodic Driving}}},
  author = {Raz, O. and Suba{\c s}{\i}, Y. and Jarzynski, C.},
  year = 2016,
  journal = {Phys. Rev. X},
  volume = {6},
  pages = {021022},
  doi = {10.1103/PhysRevX.6.021022}
}

@book{rivasOpenQuantumSystems2012,
  title = {Open {{Quantum Systems}}: {{An Introduction}}},
  author = {Rivas, {\'A}ngel and Huelga, Susana F.},
  year = 2012,
  series = {{{SpringerBriefs}} in {{Physics}}},
  edition = {1st},
  publisher = {Springer Berlin Heidelberg},
  address = {Berlin, Heidelberg},
  doi = {10.1007/978-3-642-23354-8},
  isbn = {978-3-642-23353-1 978-3-642-23354-8}
}

@article{rodriguesThermodynamicsWeaklyCoherent2019,
  title = {Thermodynamics of {{Weakly Coherent Collisional Models}}},
  author = {Rodrigues, Franklin L. S. and De Chiara, Gabriele and Paternostro, Mauro and Landi, Gabriel T.},
  year = 2019,
  journal = {Phys. Rev. Lett.},
  volume = {123},
  pages = {140601},
  doi = {10.1103/PhysRevLett.123.140601}
}

@article{satoreMarkovTypeStateModels2025,
  title = {Markov-{{Type State Models}} to {{Describe Non-Markovian Dynamics}}},
  author = {Satore, Sofia and Teichmann, Franziska and Stock, Gerhard},
  year = 2025,
  journal = {J. Chem. Theory Comput.},
  volume = {21},
  pages = {2757--2765},
  doi = {10.1021/acs.jctc.4c01630}
}

@article{schillingCoarsegrainedModellingOut2022,
  title = {Coarse-Grained Modelling out of Equilibrium},
  author = {Schilling, Tanja},
  year = 2022,
  month = aug,
  journal = {Phys. Rep.},
  volume = {972},
  pages = {1--45},
  doi = {10.1016/j.physrep.2022.04.006}
}

@article{schnellThereFloquetLindbladian2020,
  title = {Is There a {{Floquet Lindbladian}}?},
  author = {Schnell, Alexander and Eckardt, Andr{\'e} and Denisov, Sergey},
  year = 2020,
  journal = {Phys. Rev. B},
  volume = {101},
  pages = {100301},
  doi = {10.1103/PhysRevB.101.100301}
}

@article{seahCollisionalQuantumThermometry2019,
  title = {Collisional {{Quantum Thermometry}}},
  author = {Seah, Stella and Nimmrichter, Stefan and Grimmer, Daniel and Santos, Jader P. and Scarani, Valerio and Landi, Gabriel T.},
  year = 2019,
  journal = {Phys. Rev. Lett.},
  volume = {123},
  pages = {180602},
  doi = {10.1103/PhysRevLett.123.180602}
}

@article{seiferthCoarseGrainingBiochemical2020,
  title = {Coarse Graining of Biochemical Systems Described by Discrete Stochastic Dynamics},
  author = {Seiferth, David and Sollich, Peter and Klumpp, Stefan},
  year = 2020,
  journal = {Phys. Rev. E},
  volume = {102},
  pages = {062149},
  doi = {10.1103/PhysRevE.102.062149}
}

@article{seifertStochasticThermodynamicsFluctuation2012,
  title = {Stochastic Thermodynamics, Fluctuation Theorems and Molecular Machines},
  author = {Seifert, Udo},
  year = 2012,
  journal = {Rep. Prog. Phys.},
  volume = {75},
  pages = {126001},
  doi = {10.1088/0034-4885/75/12/126001}
}

@article{sinitsynUniversalGeometricTheory2007,
  title = {Universal {{Geometric Theory}} of {{Mesoscopic Stochastic Pumps}} and {{Reversible Ratchets}}},
  author = {Sinitsyn, N. A. and Nemenman, Ilya},
  year = 2007,
  journal = {Phys. Rev. Lett.},
  volume = {99},
  pages = {220408},
  doi = {10.1103/PhysRevLett.99.220408}
}

@article{strasbergNonMarkovianityNegativeEntropy2019,
  title = {Non-{{Markovianity}} and Negative Entropy Production Rates},
  author = {Strasberg, Philipp and Esposito, Massimiliano},
  year = 2019,
  journal = {Phys. Rev. E},
  volume = {99},
  pages = {012120},
  doi = {10.1103/PhysRevE.99.012120}
}

@article{strasbergQuantumInformationThermodynamics2017,
  title = {Quantum and {{Information Thermodynamics}}: {{A Unifying Framework Based}} on {{Repeated Interactions}}},
  author = {Strasberg, Philipp and Schaller, Gernot and Brandes, Tobias and Esposito, Massimiliano},
  year = 2017,
  journal = {Phys. Rev. X},
  volume = {7},
  pages = {021003},
  doi = {10.1103/PhysRevX.7.021003}
}

@article{strasbergRepeatedInteractionsQuantum2019,
  title = {Repeated {{Interactions}} and {{Quantum Stochastic Thermodynamics}} at {{Strong Coupling}}},
  author = {Strasberg, Philipp},
  year = 2019,
  journal = {Phys. Rev. Lett.},
  volume = {123},
  pages = {180604},
  doi = {10.1103/PhysRevLett.123.180604}
}

@article{tellobreuerBenchmarkingQuantumMaster2024,
  title = {Benchmarking Quantum Master Equations beyond Ultraweak Coupling},
  author = {Tello Breuer, Camilo Santiago and Becker, Tobias and Eckardt, Andr{\'e}},
  year = 2024,
  journal = {Phys. Rev. B},
  volume = {110},
  pages = {064319},
  doi = {10.1103/PhysRevB.110.064319}
}

@article{tepperAccurateMemoryKernel2024,
  title = {Accurate {{Memory Kernel Extraction}} from {{Discretized Time-Series Data}}},
  author = {Tepper, Lucas and Dalton, Benjamin and Netz, Roland R.},
  year = 2024,
  journal = {J. Chem. Theory Comput.},
  volume = {20},
  pages = {3061--3068},
  doi = {10.1021/acs.jctc.3c01289},
  copyright = {https://creativecommons.org/licenses/by/4.0/}
}

@article{vollmarModelfreeInferenceMemory2024,
  title = {Model-Free Inference of Memory in Conformational Dynamics of a Multi-Domain Protein},
  author = {Vollmar, Leonie and Bebon, Rick and Schimpf, Julia and Flietel, Bastian and Celiksoy, Sirin and S{\"o}nnichsen, Carsten and Godec, Alja{\v z} and Hugel, Thorsten},
  year = 2024,
  journal = {J. Phys. A: Math. Theor.},
  volume = {57},
  pages = {365001},
  doi = {10.1088/1751-8121/ad6d1e}
}

@article{vonkeyserlingkOperatorHydrodynamicsOTOCs2018,
  title = {Operator {{Hydrodynamics}}, {{OTOCs}}, and {{Entanglement Growth}} in {{Systems}} without {{Conservation Laws}}},
  author = {Von Keyserlingk, C. W. and Rakovszky, Tibor and Pollmann, Frank and Sondhi, S. L.},
  year = 2018,
  journal = {Phys. Rev. X},
  volume = {8},
  pages = {021013},
  doi = {10.1103/PhysRevX.8.021013}
}

@article{vroylandtLikelihoodbasedNonMarkovianModels2022,
  title = {Likelihood-Based Non-{{Markovian}} Models from Molecular Dynamics},
  author = {Vroylandt, Hadrien and Gouden{\`e}ge, Ludovic and Monmarch{\'e}, Pierre and Pietrucci, Fabio and Rotenberg, Benjamin},
  year = 2022,
  month = mar,
  journal = {Proc. Natl. Acad. Sci. U.S.A.},
  volume = {119},
  pages = {e2117586119},
  doi = {10.1073/pnas.2117586119}
}

@article{vrugtProjectionOperatorsStatistical2020,
  title = {Projection Operators in Statistical Mechanics: A Pedagogical Approach},
  shorttitle = {Projection Operators in Statistical Mechanics},
  author = {Vrugt, Michael Te and Wittkowski, Raphael},
  year = 2020,
  journal = {Eur. J. Phys.},
  volume = {41},
  pages = {045101},
  doi = {10.1088/1361-6404/ab8e28}
}

@article{wachtelThermodynamicallyConsistentCoarse2018,
  title = {Thermodynamically Consistent Coarse Graining of Biocatalysts beyond {{Michaelis}}--{{Menten}}},
  author = {Wachtel, Artur and Rao, Riccardo and Esposito, Massimiliano},
  year = 2018,
  journal = {New J. Phys.},
  volume = {20},
  pages = {042002},
  doi = {10.1088/1367-2630/aab5c9}
}

@article{wisniewskiMemoryinducedCurrentReversal2025a,
  title = {Memory-Induced Current Reversal of {{Brownian}} Motors},
  author = {Wi{\'s}niewski, Mateusz and Spiechowicz, Jakub},
  year = 2025,
  journal = {Phys. Rev. E},
  volume = {111},
  pages = {024130},
  doi = {10.1103/PhysRevE.111.024130}
}

@article{yvinecAdiabaticReductionModel2014,
  title = {Adiabatic Reduction of a Model of Stochastic Gene Expression with Jump {{Markov}} Process},
  author = {Yvinec, Romain and Zhuge, Changjing and Lei, Jinzhi and Mackey, Michael C.},
  year = 2014,
  journal = {J. Math. Biol.},
  volume = {68},
  pages = {1051--1070},
  doi = {10.1007/s00285-013-0661-y}
}

@article{zhaoEmergenceMemoryEquilibrium2024,
  title = {Emergence of {{Memory}} in {{Equilibrium}} versus {{Nonequilibrium Systems}}},
  author = {Zhao, Xizhu and Hartich, David and Godec, Alja{\v z}},
  year = 2024,
  journal = {Phys. Rev. Lett.},
  volume = {132},
  pages = {147101},
  doi = {10.1103/PhysRevLett.132.147101}
}

@article{bucaRule54ExactlySolvableModel2021,
author = {Bu{\v{c}}a, Berislav and Klobas, Katja and Prosen, Toma{\v{z}}},
doi = {10.1088/1742-5468/ac096b},
issn = {1742-5468},
journal = {J. Stat. Mech.: Theory Exp. },
number = {7},
pages = {074001},
title = {{Rule 54: exactly solvable model of nonequilibrium statistical mechanics}},
url = {https://iopscience.iop.org/article/10.1088/1742-5468/ac096b},
volume = {2021},
year = {2021}
}

@article{loosIrreversibilityHeatAndInformation2020,
author = {Loos, Sarah A M and Klapp, Sabine H L},
doi = {10.1088/1367-2630/abcc1e},
issn = {1367-2630},
journal = {New J. Phys.},
month = {dec},
number = {12},
pages = {123051},
title = {{Irreversibility, heat and information flows induced by non-reciprocal interactions}},
url = {https://iopscience.iop.org/article/10.1088/1367-2630/abcc1e},
volume = {22},
year = {2020}
}

@article{straubeRapidOnset2020,
author = {Straube, Arthur V. and Kowalik, Bartosz G. and Netz, Roland R. and H{\"{o}}fling, Felix},
doi = {10.1038/s42005-020-0389-0},
issn = {2399-3650},
journal = {Communications Physics},
month = {jul},
number = {1},
pages = {126},
title = {{Rapid onset of molecular friction in liquids bridging between the atomistic and hydrodynamic pictures}},
url = {https://www.nature.com/articles/s42005-020-0389-0},
volume = {3},
year = {2020}
}

@article{coghiCurrentFluctuations2024,
  title={Current fluctuations of a self-interacting diffusion on a ring},
  author={Coghi, Francesco},
  journal={J. Phys. A: Math. Theor.},
  volume={58},
  number={1},
  pages={015002},
  year={2024},
  url = {https://iopscience.iop.org/article/10.1088/1751-8121/ad9788/meta},
  publisher={IOP Publishing}
}

@article{lesnickiMolecularHydrodynamics2016,
  title = {Molecular Hydrodynamics from Memory Kernels},
  author = {Lesnicki, Dominika and Vuilleumier, Rodolphe and Carof, Antoine and Rotenberg, Benjamin},
  journal = {Phys. Rev. Lett.},
  volume = {116},
  issue = {14},
  pages = {147804},
  numpages = {5},
  year = {2016},
  month = {Apr},
  publisher = {American Physical Society},
  doi = {10.1103/PhysRevLett.116.147804},
  url = {https://link.aps.org/doi/10.1103/PhysRevLett.116.147804}
}

@article{jungIterativeReconstruction2017,
author = {Jung, Gerhard and Hanke, Martin and Schmid, Friederike},
title = {Iterative Reconstruction of Memory Kernels},
journal = {J. Chem. Theory Comput.},
volume = {13},
number = {6},
pages = {2481-2488},
year = {2017},
doi = {10.1021/acs.jctc.7b00274},
URL = {https://doi.org/10.1021/acs.jctc.7b00274}
}

@article{meyerNumericalProcedure2020,
author = {Meyer, Hugues and Wolf, Steffen and Stock, Gerhard and Schilling, Tanja},
title = {A Numerical Procedure to Evaluate Memory Effects in Non-Equilibrium Coarse-Grained Models},
journal = {Adv. Theor. Simul.},
volume = {4},
number = {4},
pages = {2000197},
doi = {https://doi.org/10.1002/adts.202000197},
year = {2021}
}

\let\addcontentsline\oldaddcontentsline

\clearpage

\onecolumngrid

\begin{center}
	{\large \bf Weak-Memory Dynamics in Discrete Time: Supplemental Material} \\
	\vspace{2ex} {\normalsize Hugues Meyer and Kay Brandner}\\
	\vspace{2ex} {\it School of Physics and Astronomy, University of Nottingham, 
		Nottingham NG7 2RD, United Kingdom and Centre for the Mathematics and Theoretical
		Physics of Quantum Non-Equilibrium Systems,\\ University of Nottingham, Nottingham NG7 2RD,
		United Kingdom}\\
	\vspace{2ex}
\end{center}

\tableofcontents

\newpage

\section{Master Theorem}

This section provides a detailed proof of our master theorem, which can be formulated as follows.\\

\noindent
\textbf{Theorem.}
\emph{
	Let $X_0\in\bC^N$ be given and denote by $(X_n)_{n\geq 0}\in\bC^N$ the sequence generated by the recurrence relation 
	\begin{equation}\label{eq:mt:NLEEq}
	X_{n+1} = \sV X_n + \sum_{m=1}^n \sK_m X_{n-m},
	\end{equation}
	where $\sV\in\bC^{N\times N}$ and $(\sK_n)_{n\geq 1}\in\bC^{N\times N}$. 
	Assume there exist non-negative constants $v,k,M$ such that \footnote{We note that $M^\ast = M^\ast(k,v)$ is a smooth function in $k$ and $v$. 
		Furthermore, the bound $M<M^\ast$ implies $\varepsilon= 4M/(k-v)^2 <1$ in both cases, $v+k<2$ and $v+k\geq 2$.}
	\begin{equation}\label{eq:mt:WMC2}
	k < \min\{v,1\}, \qquad 
	M < M^\ast = \begin{cases}(v-k)^2/4, & v+k <2 \\ (1-k)(v-1), & v+k \geq 2\end{cases}
	\end{equation}
	and
	\begin{equation}\label{eq:mt:WMC1}
	\nrm{\sV^{-1}}\leq 1/v, \qquad \nrm{\sK_n}\leq Mk^{n-1},
	\end{equation}
	where $\nrm{\cdot}$ denotes the spectral norm, then the following results hold true.\\
	\textbf{(i)} There exists a unique pair of a non-singular effective generator $\sG\in\bC^{N\times N}$ and a non-singular slippage matrix $\sD\in\bC^{N\times N}$ such that, for any $X_0$, 
	\begin{equation}\label{eq:mt:SMAsm}
	\lim_{n\rightarrow\infty}\sG^{-n}X_n = \sD X_0,
	\end{equation}
	and the long-time approximation $Y_n=\sG^n\sD X_0$ satisfies 
	\begin{equation}\label{eq:mt:LTAErr}
	\abs{X_n - Y_n}\leq\frac{\zeta-k}{\eta-\zeta}\abs{X_0}\zeta^n,
	\end{equation}
	where $\abs{\cdot}$ indicates the Euclidean norm and 
	\begin{equation}\label{eq:mt:ZetEta}
	\zeta = \frac{v+k-\sqrt{(v-k)^2-4M}}{2}<1, \qquad
	\eta  = \frac{v+k+\sqrt{(v-k)^2-4M}}{2}.
	\end{equation}
	\textbf{(ii)} The slippage matrix admits the explicit expression  
	\begin{equation}\label{eq:mt:SMDef}
	\sD^{-1} = 1+ \sum_{m=1}^\infty\sum_{n=1}^\infty \sH^{-m}\sK_{m+n-1}\sG^{-n},
	\end{equation}
	where $\sG$ and $\sH$ are the unique solutions of the fixed-point equations 
	\begin{align}
	\label{eq:mt:GFPEq}
	\sG &= \sV + \sum_{n=1}^\infty \sK_n\sG^{-n} = T(\sG), \\
	\label{eq:mt:HFPEq}
	\sH &= \sV + \sum_{n=1}^\infty \sH^{-n}\sK_n = U(\sH)
	\end{align}
	in the set $B= \{\sX\in\bC^{N\times N} \; : \; \nrm{\sX^{-1}}\leq 1/\eta\}$.\\
}

The proof of this theorem follows the approach developed in Refs.~[59, 60] in the main text  for continuous-time dynamics.
We proceed in four main steps.
After providing a heuristic derivation of the fixed-point equations~\eqref{eq:mt:GFPEq} and \eqref{eq:mt:HFPEq}, we show that these equations have a unique solution in~$B$.
In the second step, we demonstrate that the thus identified effective generator~$\sG$ yields a faithful long-time approximation, in the sense that $\lim_{n\rightarrow\infty}\sG^{-n}X_n = \sD X_0$ for any~$X_0$, where~$\sD$ is the slippage matrix defined in Eq.~\eqref{eq:mt:SMDef}.
The third and fourth steps of our proof establish the bound~\eqref{eq:mt:LTAErr} and the uniqueness of the effective generator and the associated slippage matrix, respectively.
In the final part of this section, we derive the properties of the memory function and the perturbation scheme discussed towards the end of the main text.

\subsection{Effective Generator}\label{sec:effective_propagator}

To analyze the asymptotic structure of the solutions of Eq.~\eqref{eq:mt:NLEEq}, we define the propagator $(\sZ_n)_{n\geq 0}\in\bC^{N\times N}$ such that $X_n = \sZ_n X_0$ for any $X_0$. 
This sequence satisfies the recursion relation 
\begin{equation}\label{eq:eg:FEEq}
\sZ_{n+1} = \sV\sZ_n + \sum_{m=1}^n\sK_m\sZ_{n-m}, \qquad	
\sZ_0 = \mathsf{1},
\end{equation}
which is equivalent to the adjoint relation 
\begin{equation}\label{eq:eg:BEEq}
\sZ_{n+1} = \sZ_n\sV + \sum_{m=1}^n\sZ_{n-m}\sK_m, \qquad 
\sZ_0 = \mathsf{1},
\end{equation}
as can be verified by noting that both define the same generating function $\hat{\sZ}_s = \sum_{n=0}^\infty \sZ_n s^n=[1-s\sV -s\hat{\sK}_s]^{-1}$, where $\hat{\sK}_s=\sum_{n=1}^\infty\sK_n s^n$.  
Upon inserting the ansatz $\sZ_n = \sG^n\sD + \sR_n$, where $\sG$ and $\sD$ are non-singular, into Eq.~\eqref{eq:eg:FEEq}, we obtain the identity 
\begin{equation}
\sG = \sV + \sum_{m=1}^n \sK_m \sG^{-m} + 
\biggl(\sV\sR_n + \sum_{m=1}^n\sK_m\sR_{n-m} - \sR_{n+1}\biggr)\sD^{-1}\sG^{-n},
\end{equation}
which reduces to the fixed-point equation \eqref{eq:mt:GFPEq} for $n\rightarrow\infty$, provided that the remainder $\sR_n$ decays sufficiently fast in this limit. 
As we shall see in the following, this assumption does not imply additional any conditions; instead, it is fulfilled whenever the weak-memory conditions \eqref{eq:mt:WMC1} and \eqref{eq:mt:WMC2} are satisfied.
The fixed-point equation \eqref{eq:mt:HFPEq} for the adjoint effective generator $\sH$ can be motivated analogously, using the ansatz $\sZ_n = \sD\sH^n + \sS_n$ and the adjoint recurrence relation \eqref{eq:eg:BEEq}. 

Banach's fixed-point theorem makes it possible to prove that these equations have unique solutions in the set $B$ (see Ref.[62] of the main text). 
To this end, we first observe that,  for any non-singular $\sX,\sY\in\bC^{N\times N}$, we have\footnote{This result follows from the bound $\sigma_{i+j-1}(\sX + \sY) \leq \sigma_i(\sX) + \sigma_j(\sY)$, which applies to the ordered singular values $\sigma_1 (\cdot) \geq \cdots \geq \sigma_N(\cdot)$ of any matrices $\sX,\sY\in\bC^{N\times N}$, and implies $\sigma_N(\sX) \geq \sigma_N(\sY) - \sigma_1(\sX- \sY)$ (see Ref. [63] of the main text). Since $\sigma_1(\sX) = \nrm{\sX}$ and $\sigma_N(\sX) = \nrm{\sX^{-1}}^{-1}$, the latter inequality is equivalent to Eq.~\eqref{eq:eg:MatNrmInq}.} 
\begin{equation}\label{eq:eg:MatNrmInq}
\nrm{\sX^{-1}}^{-1}\geq \nrm{\sY^{-1}}^{-1} - \nrm{\sX-\sY},
\end{equation}
where $\lVert \cdot \rVert$ denotes the specatral norm.
Using this bound, it follows that, for $\sX\in B$, 
\begin{align}
\nrm{[T(\sX)]^{-1}}^{-1} & \geq \nrm{\sV^{-1}}^{-1} - \nrm{\sum_{n=1}^\infty\sK_n\sX^{-n}}\\
& \geq v - M\sum_{n=1}^\infty \eta^{-n} k^{n-1} = \eta,\nonumber 
\end{align}
where we have used that $k<\eta$, since $k<v$.  
Hence, the set $B$ is closed under the map $T$. 
Next, we define the metric $d(\sX,\sY) = \nrm{\sX-\sY}$ and note that 
\begin{equation}
d(T(\sX),T(\sY))\leq M\sum_{n=1}^\infty \nrm{\sX^{-n}-\sY^{-n}}k^{n-1}. 
\end{equation}
Thus, upon inserting the identity $\sX^{-n}-\sY^{-n} = \sum_{m=1}^n \sX^{m-n-1}(\sX-\sY)\sY^{-m}$ [64], it follows that for, $\sX,\sY\in B$, 
\begin{equation}
d(T(\sX),T(\sY)) \leq M d(\sX,\sY) \sum_{n=1}^\infty n \eta^{-n-1}k^{n-1} = qd(\sX,\sY)
\end{equation}
with $q = M/(\eta-k)^ 2 = \varepsilon/(1+\sqrt{1-\varepsilon})^2$, where $\varepsilon=4M/(v-k)^ 2$. 
Since the conditions \eqref{eq:mt:WMC2} imply $\varepsilon<1$, and thus $q<1$, we can conclude that $T$ is a contraction on the closed metric space $(B,d)$. 
Therefore, the fixed-point equation $\sG = T(\sG)$ has a unique solution in $B$. 
The same result holds true for the second fixed-point equation $\sH = U(\sH)$, as can be verified along the same lines. 

We conclude this section by noting that the two effective generators $\sG$ and $\sH$ defined above and the slippage matrix $\sD$ defined in Eq.~\eqref{eq:mt:SMDef} satisfy the intertwining relation
\begin{equation}\label{eq:eg:IntRel}
\sG\sD = \sD\sH,
\end{equation}
which can be derived in two steps. 
First, we observe that the fixed-point equations \eqref{eq:mt:GFPEq} and \eqref{eq:mt:HFPEq} imply
\begin{align}\label{eq:eg:IntRelAux}
\sD^{-1}\sG & = \sG + \sum_{m=1}^\infty\sH^{-m}\sK_m + \sum_{n=1}^\infty\sum_{m=1}^\infty
\sH^{-m}\sK_{m+n}\sG^{-n}\\
& = \sH + \sum_{n=1}^\infty\sK_n\sG^{-n}  + \sum_{n=1}^\infty\sum_{m=1}^\infty
\sH^{-m}\sK_{m+n}\sG^{-n} 
= \sH\sD^{-1}. \nonumber
\end{align}
Second, since $\sG,\sH\in B$, the bound
\begin{align}
\label{eq:eg:DBnd}
\nrm{\sD^{-1}-1} \leq \sum_{n=1}^\infty\sum_{m=1}^\infty\nrm{\sH^{-m}\sK_{m+n-1}\sG^{-n}} \leq M \sum_{n=1}^\infty\sum_{m=1}^\infty \eta^{-m-n}k^{m+n-2} = q <1 
\end{align}
holds, which shows that $\sD^{-1}$ is non-singular; hence, so is $\sD$. 
Therefore, the relations \eqref{eq:eg:IntRelAux} and \eqref{eq:eg:IntRel} are equivalent. 

\subsection{Slippage Matrix}

\newcommand{\vphp}{\vphantom{\prime}}

We take $\sG$ to be the solution of the fixed-point equation \eqref{eq:mt:GFPEq} in $B$. 
Our aim is to show that $\lim_{n\rightarrow\infty} \sG^{-n}\sZ_n = \sD$, where $\sD$ is the slippage matrix defined in Eq.~\eqref{eq:mt:SMDef}.
To this end, we define the reduced propagator $\sA_n = \sG^{-n}\sZ_n$. 
This sequence satisfies the recurrence relation 
\begin{align}
\label{eq:lta:rp:RedProRecRel}
\sA_{n+1} & = \sG^{-1}\sA_n\sV + \sum_{m=0}^{n-1}\sG^{m-n-1}\sA_m\sK_{n-m}\\
& = \sG^{-1}\sA_n\sG + \sum_{m=0}^{n-1}\sG^{m-n-1}\sA_m\sK_{n-m}
- \sum_{m=1}^\infty \sG^{-1}\sA_n\sK_m\sG^{-m},
\qquad \sA_0 = \mathsf{1}, \nonumber
\end{align}
which can be derived by first inserting the adjoint evolution equation \eqref{eq:eg:BEEq} for the propagator $\sZ_n$, and then eliminating the free generator $\sV$ using Eq.~\eqref{eq:mt:GFPEq}. 
Using this result, the asymptotic behavior of the reduced propagator can be analyzed as follows. 
We first apply the transformation $\tilde{\sA}_n = \sG^n\sA_n\sG^{-n}$ and observe that 
\begin{equation}
\tilde{\sA}_{n+1} - \tilde{\sA}_n = \sum_{m=0}^{n-1}\tilde{\sA}_m\sG^m\sK_{n-m}\sG^{-n-1}
-\sum_{m=1}^\infty\tilde{\sA}_n\sG^n\sK_m\sG^{-n-m-1}, 
\qquad \tilde{\sA}_0 = \mathsf{1}.
\end{equation}
Summing both sides of this equation over $n$ shows that   
\begin{equation}\label{eq:lta:RedProRecRel_2}
\sA_n = 1 -\sum_{m=0}^{n-1}\sum_{l=n-m-1}^\infty\sG^{m-n}\sA_m\sK_{l+1}\sG^{n-m-l-2}.
\end{equation}
Next, we define the shifted sequence $\sA'_n = \sA_n^{\vphp}-\sD$, which follows the recurrence relation
\begin{align}
\sA'_n & = 1 - \sD\biggl(1 
+ \sum_{m=1}^{n}\sum_{l=1}^\infty\sH^{-m}\sK_{m+l-1}^{\vphp}\sG^{-l}\biggr)
- \sum_{m=0}^{n-1}\sum_{l=n-m-1}^\infty\sG^{m-n}\sA'_m\sK_{l+1}^{\vphp}\sG^{n-m-l-2}\\
& = \sD\sum_{m=n}^\infty\sum_{l=m}^\infty\sH^{-m-1}\sK_{l+1}^{\vphp}\sG^{m-l-1}
- \sum_{m=0}^{n-1}\sum_{l=n-m-1}^\infty\sG^{m-n}\sA'_m\sK_{l+1}^{\vphp}\sG^{n-m-l-2}, \qquad \sA'_0 = \mathsf{1}-\sD.
\nonumber
\end{align}
Here, we have first used the intertwining relation \eqref{eq:eg:IntRel} and then inserted the definition \eqref{eq:mt:SMDef} of $\sD$. 
Since $\sG,\sH\in B$, this result leads to the recursive bound  
\begin{align}
\nrm{\sA'_n} & \leq \nrm{\sD}\sum_{m=n}^\infty\sum_{l=m}^\infty
\nrm{\sH^{-m-1}\sK_{l+1}^{\vphp}\sG^{m-l-1}}+ \sum_{m=0}^{n-1}\sum_{l=n-m-1}^\infty
\nrm{\sG^{m-n}\sA'_m\sK_{l+1}^{\vphp}\sH^{n-m-l-2}}\\
& \leq \nrm{\sD}M\sum_{m=n}^\infty\sum_{l=m}^\infty	
\eta^{-l-2}k^l
+M\sum_{m=0}^{n-1}\sum_{l=n-m-1}^\infty \nrm{\sA'_m}\eta^{-l-2}k^l \nonumber\\
& = \nrm{\sD}q\eta^{-n}k^n
+\frac{M}{\eta-k} \sum_{m=0}^{n-1}\nrm{\sA'_m}\eta^{m-n}k^{n-m-1}\nonumber
\end{align} 
with $\nrm{\sA'_0} = \nrm{\mathsf{1-\sD}} = \nrm{\sD(\sD^{-1}-\mathsf{1})}\leq \nrm{\sD}q$.
Here, we have inserted the bound \eqref{eq:eg:DBnd} with the parameter $q$ being defined as $q = M/(\eta-k)^ 2 = \varepsilon/(1+\sqrt{1-\varepsilon})^2$, where $\varepsilon=4M/(v-k)^ 2$.
These results imply
\begin{equation}
\nrm{\sA'_n}\leq \nrm{\sD}qp^n 
\quad\text{with}\quad
p = \frac{k(\eta-k)+ M}{\eta(\eta-k)} 
= \frac{k+v-(v-k)\sqrt{1-\varepsilon}}{k+v+(v-k)\sqrt{1-\varepsilon}} < 1,
\end{equation}
as can be verified by induction. 
Consequently, we have shown that 
\begin{equation}
\lim_{n\rightarrow\infty}\nrm{\sA'_n}
= \lim_{n\rightarrow\infty}\nrm{\sA_n^{\vphp}-\sD} 
= \lim_{n\rightarrow\infty}\nrm{\sG^{-n}\sZ^{\vphp}_n-\sD} = 0. 
\end{equation}
The corresponding result for the adjoint effective generator, 
\begin{equation}
\lim_{n\rightarrow\infty}\nrm{\sZ_n\sH^{-n} -\sD} = 0,
\end{equation}
can be obtained along the same lines. 

\subsection{Error Bound}

To derive the bound \eqref{eq:mt:LTAErr}, we define the memory function $\sE_n = \sG^{n+1}(\sA_{n+1} -\sA_n)$. 
This sequence satisfies the recurrence relation 
\begin{equation}
\sE_n =  - \sum_{m=n}^\infty\sK_{m+1}\sG^{n-m-1}
- \sum_{m=0}^{n-1}\sum_{l=m}^\infty\sE_{n-m-1}\sK_{l+1}\sG^{m-l-1},
\qquad \sE_0 = \sV - \sG,
\end{equation}
which follows from the recurrence relation for the reduced propagator \eqref{eq:lta:RedProRecRel_2}, and implies the recursive bound 
\begin{align}
\nrm{\sE_n} & \leq \sum_{m=n}^\infty \nrm{\sK_{m+1}\sG^{n-m-1}} 
+ \sum_{m=0}^{n-1}\sum_{l=m}^\infty\nrm{\sE_{n-m-1}\sK_{l+1}\sG^{m-l-1}}\\
& \leq M\sum_{m=n}^\infty \eta^{n-m-1}k^m + M\sum_{m=0}^{n-1}\sum_{l=m}^\infty
\nrm{\sE_{n-m-1}}\eta^{m-l-1}k^l\nonumber\\			
&=\frac{Mk^n}{\eta-k} + \frac{M}{\eta-k}\sum_{m=0}^{n-1}\nrm{\sE_{n-m-1}}k^m\nonumber
\end{align} 
with 
\begin{align}
\nrm{\sE_0} = \nrm{\sV-\sG} & \leq \sum_{n=1}^\infty\nrm{\sK_n\sG^{-n}}
\leq M \sum_{n=1}^\infty \eta^{-n}k^{n-1} 
= \frac{M}{\eta-k}=\zeta- k.
\end{align}
Here, we have used the parameters $\zeta$ and $\eta$ being defined in Eq.~\eqref{eq:mt:ZetEta}, and the fact that $(\eta-k)(\zeta-k)=M$.
Consequently, we have 
\begin{equation}\label{eq:eb:CMFBnd}
\nrm{\sE_n} \leq (\zeta-k)\zeta^n,
\end{equation}
as can again be shown by induction. 
Next, we note that, by construction, $\sA_n = 1+\sum_{m=1}^n\sG^{-m}\sE_{m-1}$. 
Thus, the slippage matrix must be given by $\sD= 1+\sum_{m=1}^\infty\sG^{-m}\sE_{m-1}$, and the error of the long-time approximation $Y_n = \sG^n\sD X_0$ of $X_n$ can be expressed as 
\begin{align}
\abs{X_n - Y_n} = \abs{\sG^n(\sA_n - \sD)X_0} 
& \leq \abs{X_0}\sum_{m=n+1}^\infty\nrm{\sG^{n-m}\sE_{m-1}}\\
& \leq (\zeta-k)\abs{X_0}\sum_{m=n+1}^\infty\eta^{n-m}\zeta^{m-1}
= \frac{\zeta-k}{\eta-\zeta}\abs{X_0}\zeta^n, \nonumber
\end{align}
where we have used that the matrix norm $\nrm{\cdot}$ and the vector norm $\abs{\cdot}$ are consistent. 

\subsection{Uniqueness}

To establish the uniqueness of the effective generator, we let $\bar{\sG}\in\bC^{N\times N}$ be a non-singular matrix such that $\sZ_n = \bar{\sG}^n\bar{\sA}_n$ and the sequence $(\bar{\sA}_n)_{n\geq 0}\in\bC^{N\times N}$ converges to a non-singular matrix $\bar{\sD}\in\bC^{N\times N}$ in the limit $n\rightarrow\infty$. 
We assume that $\bar{\sG}\neq\sG$ and $\bar{\sD}\neq\sD$. 
However, since the propagator is unique, we must still have $\bar{\sG}^n\bar{\sA}_n = \sG^n\sA_n$. 
By assumption, the sequence $\sS_n = \bar{\sG}^{-n}\sG^n=\bar{\sA}_n^{\vphantom{1}}\sA^{-1}_n$ converges to a non-singular matrix $\sS = \bar{\sD}\sD^{-1}$ for $n\rightarrow\infty$. 
Consequently, we have 
\begin{equation}
0 = \lim_{n\rightarrow\infty}\bigl(\sS_{n+1}-\sS_n\bigr) 
= \lim_{n\rightarrow\infty}\bigl(\bar{\sG}^{-1}\sS_n\sG - \sS_n\bigr)
= \bar{\sG}^{-1}\sS\sG - \sS. 
\end{equation}
Hence, $\sG$ and $\bar{\sG}=\sS\sG\sS^{-1}$ are similar to each other. 
Next, we define $\bar{Y}_n = \bar{\sG}^n\bar{\sD}X_0= \sS\sG^n\sD X_0$ and note that 
\begin{equation}\label{eq:uq:SubBnd}
\abs{X_n-\bar{Y}_n}\geq \bigl|\abs{\bar{Y}_n-Y_n}-\abs{X_n-Y_n}\bigr|,
\end{equation}
where $Y_n = \sG^n\sD X_0$ is the canonical long-time approximation that satisfies the bound \eqref{eq:mt:LTAErr}. 
Since $\sD$ is non-singular, and $\sS-1 \neq 0$ by assumption there exist an initial state $X_0\neq 0$ and a constant $C>0$ such that\footnote{To derive this result, first assume that $\sG$ has an eigenvector $G^j_1$ with corresponding eigenvalue $\gamma_j$ such that $(1-\sS)G^1_j\neq0$.
	Setting $\sD X_0 = G^1_j$ then yields $ \abs{(\sS-1)\sG^n\sD X_0}= \abs{(\sS-1)G^1_j}\abs{\gamma_j^{\vphantom{1}}}^n \geq \abs{(\sS-1)G^1_j}\eta^n$, 
	where the last inequality follows from the condition $\nrm{\sG^{-1}}\leq 1/\eta$.
	If $(\sS-1)G^1_j=0$ for all eigenvectors of $\sG$, assume that $\sG$ has a generalized eigenvector $G^2_j$ of rank $2$ such that $(\sS-1)G^2_j\neq 0$. 
	Since $\sG G^2_j = \gamma^{\vphantom{1}}_jG^2_j+G^1_j$ and $(1-\sS)G^1_j =0$, we then have $(\sS-1)\sG^n G^2_j = \abs{\gamma^{\vphantom{1}}_j}^n(\sS-1)G^2_j$, and the result \eqref{eq:uq:DevBnd} follows by setting $\sD X_0 = G^2_j$. 
	If $(\sS-1)G^1_j = (\sS-1)G^2_j =0$ for all $j$, the above argument can be repeated with generalized eigenvectors of successively higher rank. 
	Since the proper and generalized eigenvectors of $\sG$ together form a complete basis, and $\sS-1\neq 0$, there must eventually exist a $G^m_j$ such that $(\sS-1)G^m_j \neq 0$. 
}
\begin{align}
\label{eq:uq:DevBnd}
\abs{\bar{Y}_n - Y_n} = \abs{(\sS-1)\sG^n\sD X_0}  \geq C\eta^n
\end{align}
Upon recalling Eq.~\eqref{eq:uq:SubBnd} and the bound \eqref{eq:mt:LTAErr} on the error of the canonical long-time approximation $Y_n$, and using, this observation implies
\begin{equation}
\abs{X_n-\bar{Y}_n}\geq C \eta^n- \frac{\zeta-k}{\eta-\zeta}\abs{X_0}\zeta^n,
\end{equation}
where we have used that $\eta > \zeta$.
This result shows that, for any effective generator $\bar{\sG}\neq\sG$ that gives rise to a faithful long-time approximation, $\bar{Y}_n\neq Y_n$, there exist initial states $X_0$ such that 
\begin{equation}
\lim_{n\rightarrow\infty}\abs{X_n - \bar{Y}_n}\sigma^{-n} = \infty
\end{equation}
for any $\sigma<\eta$. 
That is, if $\sG$ is non-singular and the asymptotic relation \eqref{eq:mt:SMAsm} is satisfied with a non-singular slippage matrix $\sD$, then the bound \eqref{eq:mt:LTAErr} holds for any $X_0$ if and only if $\sG$ is the unique solution of the fixed-point equation \eqref{eq:mt:GFPEq} in the set $B$ defined after Eq.~\eqref{eq:mt:HFPEq}, and the slippage matrix is given by Eq.~\eqref{eq:mt:SMDef}. 
The proof of our master theorem is therefore complete. 

\subsection{Memory Function}

\newcommand{\vphx}{\vphantom{\sX}}

For $\sX\in\bC^{N\times N}$, we define the memory function $(\sE_n(\sX))_{n\geq 0}\in\bC^{N\times N}$ such that the propagator of Eqs.~\eqref{eq:eg:FEEq} and \eqref{eq:eg:BEEq} satisfies the inhomogeneous recurrence relation
\begin{equation}\label{eq:mf:MFDef}
\sZ_{n+1} = \sX\sZ_n + \sE_n(\sX), \qquad \sZ_0 = \mathsf{1}.
\end{equation}
This condition implies that $\sE_n(\sX)$ is uniquely determined by the initial-value problem\footnote{This result follows by comparing the Laplace space solutions of Eq.~\eqref{eq:mf:MFDef}, $\hat{\sZ}_s= [1-s\hat{\sX}]^{-1}(1+s\hat{\sE}_s(\sX))$, and Eqs.~\eqref{eq:eg:FEEq} and \eqref{eq:eg:BEEq}, $\hat{\sZ}_s= [1-s\sV-s\hat{\sK}_s]^{-1}$, where $\hat{\sZ}_s =\sum_{n=0}^\infty\sZ_n s^n$, $\hat{\sK}_s = \sum_{n=1}^\infty \sK_n s^n$ and $\hat{\sE}_s(\sX)= \sum_{n=0}^\infty \sE_n(\sX)s^n$.}
\begin{equation}\label{eq:mf:MFRec}
\sE_{n+1}(\sX) = \sK_{n+1} + \sE_n(\sX)\sV + \sum_{m=1}^n\sE_{n-m}(\sX)\sK_m, 
\qquad \sE_0(\sX) = \sV - \sX. 
\end{equation}
In the following, we establish two key properties of the sequence $\sE_n(\sX)$. 
First, the bound \eqref{eq:eb:CMFBnd} applies only to the canonical memory function $\sE_n(\sG) = \sE_n$. 
Second, if $\sX\neq\sG$, the asymptotic relation 
\begin{equation}
\limsup_{n\rightarrow\infty} \nrm{\sE_n(\sX)}\sigma^{-n} = \infty
\end{equation}
holds for any $\sigma<\eta$. 
Upon recalling that the bound $\nrm{\sE_n}\leq (\zeta-k)\zeta^n$ holds for the canonical memory function, and that $\eta>\zeta$, it follows that the canonical effective generator $\sG$ is unique in that it corresponds to the fastest decaying memory function. 

To derive these results, we first convert the initial-value problem \eqref{eq:mf:MFRec} into a final-value problem by introducing the sequence $\tilde{\sE}_n(\sX) = \sE_n(\sX)\sV^{-n}$, which satisfies
\begin{equation}\label{eq:mf:MFTrsRec}
\tilde{\sE}_{n+1}(\sX)-\tilde{\sE}_n(\sX) = \sK_{n+1}\sV^{-n-1}
+\sum_{m=1}^n \tilde{\sE}_{n-m}(\sX)\sV^{n-m}\sK_m\sV^{-n-1}, \qquad
\tilde{\sE}_0(\sX)=\sV -\sX.
\end{equation}
Next, for any $0<S$ and $0<\sigma<1$, we define a set of rapidly decaying sequences, 
\begin{equation}
\fB_{S,\sigma}
= \bigl\{(\sX_n)_{n\geq 0}\in\bC^{N\times N} \; : \; \nrm{\sX_n}\leq S\sigma^n\bigr\}. 
\end{equation}
If $\sE_n(\sX)\in\fB_{S,\sigma}$ with $\sigma<v$, we have $\lim_{n\rightarrow\infty}\tilde{\sE}_n(\sX) = 0$ and, upon summing over $n$, Eq.~\eqref{eq:mf:MFTrsRec} implies 
\begin{align}
\sE_n(\sX)  = -\sum_{m=n}^\infty\sK_{m+1}\sV^{n-m-1}
-\sum_{m=n}^\infty\sum_{l=1}^m\sE_{m-l}(\sX)\sK_l\sV^{n-m-1}
= \fT_n[\sE(\sX)],\
\end{align}
where $\fT$ defines a linear map on the space of matrix-valued sequences. 
We now observe that, for $\sX\in\fB_{S,\sigma}$, 
\begin{align}
\nrm{\fT_n[\sX]} & \leq \sum_{m=n}^\infty\nrm{\sK_{m+1}}\sV^{n-m-1}
+ \sum_{m=n}^\infty\sum_{l=1}^m\nrm{\sX_{m-l}\sK_l\sV^{n-m-1}}\\
& \leq M \sum_{m=n}^\infty k^{m}v^{n-m-1} + MS\sum_{m=n}^\infty\sum_{l=1}^m 
k^{l-1} v^{n-m-1}\sigma^{m-l}\nonumber\\
& =\frac{M}{\sigma-k}\biggl(\frac{S\sigma^n}{v-\sigma}-\frac{(k-\sigma+S)k^n}{v-k}\biggr),
\nonumber
\end{align}
where we have again used the assumption $\sigma<v$. 
Upon further restricting the parameters $S$ and $\sigma$ such that 
\begin{equation}\label{eq:mf:RDPCls}
S\geq \sigma -k \quad\text{and}\quad \zeta \leq \sigma < \min\{\eta,1\}, 
\end{equation}
with the parameters $\zeta$ and $\eta$ defined in Eq.~\eqref{eq:mt:ZetEta}, we obtain the bound 
\begin{equation}
\nrm{\fT_n[\sX]}\leq \frac{M S\sigma^n}{(\sigma-k)(v-\sigma)} \leq S\sigma^n. 
\end{equation}
That is, for any $S$ and $\sigma$ satisfying the conditions \eqref{eq:mf:RDPCls}, $\fT$ maps the set $\fB_{S,\sigma}$ of rapidly decaying sequences into itself.
For the next step of our derivation, we equip each set $\fB_{S,\sigma}$ with a metric,
\begin{equation}
d_\mu(\sX,\sY) = \sup_{n\geq 0}\nrm{\sX_n-\sY_n}\mu^{-n}
\quad\text{with}\quad \mu\geq \sigma. 
\end{equation}
We then have 
\begin{align}
d_\mu (\fT[\sX],\fT[\sY]) & \leq \sup_{n\geq 0} \sum_{m=n}^\infty\sum_{l=1}^m
\nrm{(\sX_{m-l} - \sY_{m-l})\sK_l\sV^{n-m-1}}\mu^{-n}\\
&\leq M d_\mu(\sX,\sY)\sup_{n\geq 0}
\sum_{m=n}^\infty\sum_{l=1}^m k^{l-1} v^{n-m-1}\mu^{m-l-n}\nonumber\\
&= M d_\mu(\sX,\sY) \sup_{n\geq 0} \frac{1}{\mu-k}
\biggl(\frac{1}{v-\mu}-\frac{k^n\mu^{-n}}{v-k}\biggr).\nonumber
\end{align}
for any $\sX_n,\sY_n\in\fB_{S,\sigma}$ and any $\mu<v$. 
Thus, if we further restrict $\mu$ such that $k<\mu<v$, it follows that 
\begin{equation}
d_\mu (\fT[\sX],\fT[\sY]) \leq q_\mu d_\mu(\sX,\sY),
\end{equation}
where the Lipschitz constant $q_\mu = M/(\mu-k)(v-\mu)$ becomes strictly smaller than $1$ if $\zeta < \mu < \eta$. 
That is, for any $S$ and $\sigma$ satisfying the conditions \eqref{eq:mf:RDPCls}, there exists a $\mu\geq \sigma$ such that the map $\fT$ becomes a contraction on $\fB_{S,\sigma}$ with respect to the metric $d_\mu$. 
Since $(d_\mu, \fB_{S,\sigma})$ is a complete metric space, it follows from Banach's fixed-point theorem that any admissible set $\fB_{S,\sigma}$ contains exactly one solution of the fixed-point equation $\fT_n[\sX] = \sX_n$ [62]. 

This result has two main implications for our theory. 
First, the set $\fB_{k-\zeta,\zeta}$ contains only the canonical memory function $\sE_n = \sE_n(\sG)$. 
Second, since the correspondence between $\sX$ and $\sE_n(\sX)$ is one-to-one, the bound $\nrm{\sE_n(\sX)}\leq S\sigma^n$, must be violated for any $\sX\neq\sG$ as long as $S$ and $\sigma$ satisfy the conditions \eqref{eq:mf:RDPCls}. 
That is, there must exist an $n\geq 0$ such that 
\begin{equation}
\nrm{\sE_n(\sX)} > S\sigma^n. 
\end{equation}
Since $S$ can be chosen arbitrary large, and $\nrm{\sE_n}<\infty$ for any finite $n$, this condition can only be met if
\begin{equation}
\limsup_{n\rightarrow\infty} \nrm{\sE_n(\sX)}\sigma^{-n} = \infty
\end{equation}
for any admissible $\sigma<\min\{\eta,1\}$. 

We conclude this section by noting that the mathematical structure developed above is useful for perturbative calculations. 
Specifically, if an initial guess $\sE^0_n\in\fB_{S,\sigma}$ for the canonical memory function is know, where $S$ and $\sigma$ satisfy the conditions \eqref{eq:mf:RDPCls}, then $\sE_n^{\vphantom{\ell}}=\lim_{\ell\rightarrow\infty}\sE^\ell_n$ can be arbitrary well approximated by iterating the map $\fT$, that is, by evaluating $\sE_n^{\ell +1} = \fT_n^{\vphantom{\ell}}[\sE^\ell_{\vphantom{n}}]$. 
Once the memory function has been obtained to sufficient accuracy, the corresponding effective generator and slippage matrix can be recovered through the relations 
\begin{equation}
\sG = \sV -\sE_0 \quad\text{and}\quad \sD = 1 + \sum_{n=1}^\infty\sG^{-n}\sE_{n-1}. 
\end{equation}
In particular, for $\sE_n^0=0$, we obtain a uniformly convergent series expansion $\sE_n^{\vphantom{\ell}} = \sum_{\ell=1}^\infty \sE^{(\ell)}_n$ of the canonical memory function in the memory strength, where 
\begin{equation}
\sE_n^{(1)} = -\sum_{m=n}^\infty \sK_{m+1}^{\vphantom{(\ell)}}\sV^{n-m-1}
\quad\text{and}\quad
\sE^{(\ell+1)}_n = - \sum_{m=n}^\infty\sum_{l=1}^m \sE_{m-l}^{(\ell)}\sK_l^{\vphantom{(\ell)}}\sV^{n-m-1}. 
\end{equation}
Hence, $\sE^{(\ell)}_n$ is of order $\ell$ in the memory kernel, and thus of order $\ell$ in the coupling strength $M$, which enters the weak-memory conditions \eqref{eq:mt:WMC1} through the bound $\nrm{\sK_n}\leq Mk^{n-1}$.

\newpage
\section{Application 1: Mesoscopic Charge Pump}

This section contains details on the stochastic pump model presented in the main text. 
We first provide a brief recap the projection operator formalism used to derive the equation of motion for the lumped state, and then move on to the specifics of the model discussed in the main text. 

\subsection{Projection Operators}

We consider a system with a discrete space of microstates $\mathbb{M}=\{1,\dots,M\}$, whose state is described by the probability vector $P_n =[P^1_n,\dots, P^M_n]^\sT$. 
This vector evolves according to the discrete-time master equation 
\begin{equation}
\label{eq:markov_jump}
P_{n+1} = \sL P_{n},
\end{equation}
where $\sL$ is a stochastic matrix, i.e., $(\sL)_{ij}\geq 0$ and $\sum_j (\sL)_{ij} =1$. 
Without loss of generality, we assume that the microstates are arranged in blocks of $N$ disjoint mesostates $\mathbb{M}_\alpha\subset\mathbb{M}$, each of which contains $M_\alpha$ microstates.
The occupation probabilities of these mesostates are collected in the lumped state vector $X_n = [X^1_n,\dots,X^N_n]^\sT = \sM P_n$. 
Here, the matrix 
\begin{equation}\label{eq:sp:DTME}
\sM = \bigoplus_{\alpha=1}^N (1_\alpha)^\sT
\end{equation}
projects the space of microstates onto the space of mesostates, where $1_{\alpha}$ denotes the all-ones vector of size $M_\alpha$;
that is, $X^\alpha_n = \sum_{i\in\mathbb{M}_\alpha}P^i_n$. 

To derive an equation of motion for $X_n$, we fix a reference probability vector $R_\alpha^{\vphantom{M_\alpha}}=[R^{1\vphantom{M_\alpha}}_\alpha,\dots,R^{M_\alpha}_\alpha]^\sT$ for every mesostate and define the matrix 
\begin{equation}
\sM^\ast = \bigoplus_{\alpha=1}^N R_\alpha
\end{equation}
such that $\sM\sM^\ast=\mathsf{1}$ is the identity matrix of size $N$ and $\sM^\ast\sM=\sP$ is a projection matrix of size $M$. 
Upon defining the relevant and irrelevant parts of the microscopic state vector as $P_n^\parallel = \sP P_n$ and $P_n^\perp = \sQ P_n$, respectively, where $\sQ = 1-\sP$, Eq.~\eqref{eq:sp:DTME} can be rewritten as a set of two coupled recursion relations, 
\begin{align}
\label{eq:sp:xnpar}
P_{n+1}^\parallel & = \sP\sL P_n^{\parallel} + \sP\sL P_n^\perp, \\
\label{eq:sp:xnperp}
P_{n+1}^\perp & = \sQ\sL P_n^{\parallel} + \sQ\sL P_n^\perp.
\end{align}
Recursively substituting the second of these equations into the first yields the closed evolution equation 
\begin{equation}
\label{eq:xnpar_2}
P_{n+1}^\parallel = \sP\sL P_n^{\parallel} + \sP\sL \sum_{m=1}^{n} (\sQ\sL)^{m} P_{n-m}^\parallel + \sP\sL (\sQ\sL)^n P^\perp_0
\end{equation}
for $P^\parallel_n$, which, upon recalling that $\sM P^\parallel_n = \sM P_n = X_n$, can be reduced to the lower-dimensional recurrence relation
\begin{equation}\label{eq:sp:NLEEQ}
X_{n+1} = \sV X_n +  \sum_{m=1}^{n} \sK_m X_{n-m} + x_n
\quad\text{with}\quad
\sV = \sM\sL\sM^* \quad\text{and}\quad
\sK_{n} = \sM\sL(\sQ\sL)^{n}\sM^*. 
\end{equation}
The inhomogeneous term $x_n = \sM\sL(\sQ\sL)^n P^\perp_0$ vanishes for $P^\perp_0 =0$, that is, if the reference probability vectors $R_\alpha$ reflect the initial occupation probabilities of the microstates in the individual mesostates.  

\subsection{Minimal Model}

We now turn to the mesoscopic charge pump discussed in the main text.  
In continuous time, the device is described by the master equation 
\begin{equation}
\dot{p}_t = \sW_t p_t,
\end{equation}
where $p_t = [p^{\circ\circ}_t, p^{\bullet\circ}_t, p^{\circ\bullet}_t]^\sT$ denotes the microscopic probability vector and the time-dependent rate matrix is defined as 
\begin{equation}
\sW_t = \left\{
\begin{array}{llll}
\sW_+, & 0      \!\! & \leq \;\; t\!\! \mod\tau \;\; < &\tau_+\\
\sW_0, & \tau_+ \!\! & \leq \;\; t\!\! \mod\tau \;\; < &\tau_+ + \tau_0\\
\sW_-, & \tau_+ +\tau_0 \!\! & \leq \;\; t\!\! \mod\tau \;\;< & \tau_++\tau_0+\tau_- = \tau 
\end{array}
\right.
\end{equation}
with the rate matrices 
\begin{equation}
\sW_+= \left[
\begin{array}{rrr}
- w_+ & 0 & 0\\
w_+ & 0 & 0\\
0   & 0 & 0
\end{array}
\right], \qquad
\sW_0= \left[
\begin{array}{rrr}
0 & 0 & 0\\
0 & -w_0 &  w_0\\
0 &  w_0 & -w_0
\end{array}
\right], \qquad
\sW_-= \left[
\begin{array}{rrr}
0 & 0 & w_-\\
0 & 0 & 0\\
0 & 0 & -w_-
\end{array}
\right]
\end{equation}
representing the three strokes of the operation cycle. 
Thus, the transition matrix that describes the evolution of the stroboscopic probability vector $P_n = p_{n\tau} = [P^{\circ\circ}_n, P^{\bullet\circ}_n, P^{\circ\bullet}_n]^\trans$ takes the form $\sL = e^{\sW_-\tau_-}e^{\sW_0\tau_0}e^{\sW_+\tau_+} = \sL_-\sL_0\sL_+$, and the parameters introduced in the main text are given by $L_\pm= 1 -e^{-w_\pm\tau_\pm}$ and $L_0 = (1-e^{-2w_0\tau_0})/2$. 
We now fix the reference probability vectors $R_\circ = [1]$ and $R_\bullet = [1/2,1/2]^\trans$ for the two mesostates $\circ$ and $\bullet$, which correspond to the conductor being neutral and occupied by a single charge, respectively. 
Using the fomalism outlined in the previous section, it is then straightforward to derive the expressions 
\begin{equation}
\sV = 1+ L_+(1-L_0L_-)\sJ_1 + \frac{L_-}{2}\sJ_2, \qquad
\sK_n = \frac{L_+L_-(1-2L_0+2L_-)k^n}{2-L_-}\sJ_1 
- \frac{L_-^2k^n}{2(2-L_-)}\sJ_2
\end{equation}
for the free generator and the memory kernel, where $k=(1-2L_0)(2-L_-)/2$ and we have introduced the matrices
\begin{equation}
\sJ_1 = \left[\begin{array}{rr}
-1 & 0 \\ 1 & 0
\end{array}
\right], \qquad
\sJ_2 = \left[\begin{array}{rr}
0 & 1 \\ 0 & -1
\end{array}
\right], 
\end{equation}  
which satisfy the multiplication rules $\sJ_1\sJ_2= -\sJ_2$ and $\sJ_2\sJ_1= -\sJ_1$.
For simplicity, we assume that $P^{\circ\bullet}_0 = P^{\bullet\circ}_0$ such that the inhomogeneous term in the effective evolution equation \eqref{eq:sp:NLEEQ} vanishes. 
Since the memory kernel has the simple structure $\sK_n = k^{n-1}\sK_1$, the non-linear fixed-point equations \eqref{eq:GFPE} and \eqref{eq:HFPE} of the main text reduce to the quadratic matrix equations 
\begin{equation}
\sG^2 -(k+\sV)\sG + k\sV - \sK_1 = 0, \qquad
\sH^2 -\sH(k+\sV) + k\sV - \sK_1 = 0. 
\end{equation}
The effective generator $\sG$ and its adjoint counterpart $\sH$ can therefore be obtained by setting $\sG = 1 + g_1 \sJ_1 + g_2 \sJ_2$ and $\sH = 1 + h_1 \sJ_1 + h_2 \sJ_2$ and solving the resulting quadratic equations for the parameters $g_1, g_2$ and $h_1, h_2$. 
The slippage matrix $\sD$ then follows from Eq.~\eqref{eq:SMDef} of the main text.

\section{Application 2: Collisional Model}

This section provides further details on the quantum collisional model discussed in the main text, following Ref.~[78] of the main text. 
A general collisional model consists of a system of interest and $N$ ancillas.   
The initial density matrix of this setup is given by 
\begin{equation}
\sigma_0 = \rho_0 \otimes \xi^1 \otimes \cdots \otimes \xi^N,
\end{equation}
where $\rho_0$ and $\xi^{1},\dots,\xi^N$ are the initial states of the system and the ancillas, respectively. 
Dynamics are generated by coupling the ancillas one by one to the system. 
To introduce memory effects, we further include interactions between consecutive ancillas. 
The state $\sigma_n$ of the full system after $n\leq N$ collisions then satisfies the recursion relation 
\begin{equation}
\sigma_{n+1}   = \mU_{n+1} \mQ_{n,n+1}\sigma_n,
\end{equation}
where the bipartite maps $\mU_i$ and $\mQ_{ij}$ describe interactions between the system proper and the $i^\mathrm{th}$ ancilla, and between the $i^\mathrm{th}$ and the $j^\mathrm{th}$ ancilla, respectively.  
Here, we use the convention $\mQ_{10}=\mathcal{I}$. 

Deriving a closed equation of motion for the state $\rho_n = \tr_A[\sigma_n]$ of the system, where $\tr_A[\circ]$ indicates the partial trace over the joint Hilbert space of the ancillas, is in general not straightforward. 
For simplicity, we therefore focus on a specific model, where both the system and the ancillas are qubits. 
System-ancilla and ancilla-ancilla interactions are described by the unitary and incoherent partial swap maps
\begin{equation}
\mU_i\circ = \sU_i^{\vphantom{\dagger}}\circ\sU_i^\dagger, \qquad
\mQ_{ij}\circ = (1-k)\circ+k\sS_{ij}\circ\sS_{ij}
\end{equation}
with $\sU_i=\exp[-\mathrm{i}\theta\sS_{Si}]=\sqrt{1-u}-\mathrm{i}\sqrt{u}\sS_{Si}$, where $u=\sin^ 2[\theta]$ and $0\leq \theta\leq\pi/2$. 
Here, $0<u,k<1$ are the swap probabilities and the Hermitian swap matrices are defined such that $\sS_{ij}\ket{\alpha_i,\beta_j}=\ket{\beta_i,\alpha_j}$, where $i,j=S,1,\dots,N$ and $\alpha,\beta =0,1$; the states $\ket{0_S}, \ket{1_S}$ and $\ket{0_i}, \ket{1_i}$ form the canonical bases of the system and the $i^\mathrm{th}$ ancilla, respectively. 
Provided that all ancillas are initially prepared in the same state, $\xi^i = \xi^1$, one can now show that the state of the system follows the evolution equation 
\begin{equation}\label{eq:cm:NLEEQ}
\rho_{n+1} = \mV\rho_n + \sum_{m=1}^n\mK_m\rho_{n-m}+\mF_{n+1}\rho_0,
\end{equation}
where $\mV=\mK_0$, $\mK_n= (1/k -1)\mF_{n+1}$, and $\mF_n\circ=k^n\tr_1^{\vphantom{n}}[\mU_1^n(\circ\otimes\xi^1)]$, for details see Ref.~[78] of the main text.

This recurrence relation formally has the same structure as Eq.~\eqref{eq:mt:NLEEq}, for which our general theory was developed, up to an inhomogeneous term, which can be accommodated as follows. 
We first note that the solution of Eq.~\eqref{eq:cm:NLEEQ} is given by the variation of parameters formula (see Ref. [65] of the main text)
\begin{equation}
\rho_n = \mZ_n \rho_0 + \sum_{m=1}^n\mZ_{n-m}\mF_m\rho_0,
\end{equation}
where $\mZ_n$ denotes the free propagator, which satisfies 
\begin{equation}
\mZ_{n+1} = \mV\mZ_n + \sum_{m=1}^n\mK_m\mZ_{n-m}, \qquad \mZ_0=\mathcal{I}. 
\end{equation}
Thus, upon inserting the long-time approximation $\mY_n = \mG^n \mD$ of the propagator, we obtain the expression 
\begin{equation}
\chi_n = \mG^n\mD\rho_0 + \sum_{m=1}^n\mG^{n-m}\mD\mF_m\rho_0
\end{equation}
for the long-time approximation $\chi_n$ of $\rho_n$, where $\mG$ is the effective generator and $\mD$ the slippage map.  
The error of this approximation is subject to the bound 
\begin{align}
\abs{\rho_n-\chi_n} & \leq \abs{\rho_n - \mG^n\mD\rho_0} 
+ \sum_{m=1}^n\abs{\mZ_{n-m}\mF_m\rho_0-\mG^{m-n}\mD\mF_m\rho_0}\\
& \leq \frac{\zeta-k}{\eta-\zeta}\abs{\rho_0}\zeta^n + \frac{\zeta-k}{\eta-\zeta}\abs{\rho_0}\zeta^n\sum_{m=1}^n\nrm{\mF_m}\zeta^{-m}.\nonumber
\end{align}
Here, we have used that, according to our master theorem, $\nrm{\mZ_n-\mG^n\mD}\leq (\zeta-k)\zeta^n/(\eta-\zeta)$ with $\zeta$ and $\eta$ being defined in Eq.~\eqref{eq:mt:ZetEta}, and the primary parameters $v,k,M$ must be determined such that the bounds $\nrm{\mV^{-1}}\leq 1/v$ and $\nrm{\mK_n}\leq M k^{n-1}$ are satisfied.

Throughout this section, we implicitly treat density matrices as vectors and quantum maps as matrices. 
These pictures are essentially equivalent for quantum systems with a finite-dimensional Hilbert space, whose density matrices can be represented as finite linear combinations of orthonormal basis vectors in the corresponding operator space. 
For qubits, the canonical operator basis is given by $\bigl\{\mathsf{1}/\sqrt{2},\sigma_x/\sqrt{2},\sigma_y/\sqrt{2},\sigma_z/\sqrt{2}\bigr\}$, where $\sigma_{x,y,z}$ are the usual Pauli matrices. 
In this basis, the map $\mF_n$, which defines both the free generator $\mV$ and the memory kernel $\mK_n$ of the toy model discussed above, takes the matrix form 
\begin{equation}
\mF_n\equiv \frac{k^n}{2}\left[   
\begin{array}{cccc}
2         &           0           &           0           & 0\\
0          & 2\cos^2[n\theta]  & (2\kappa-1)\sin[2n\theta] &   0\\
0          & (1-2\kappa)\sin[2n\theta] & 2\cos^2[n\theta]  &   0\\
(2-4\kappa)\sin^ 2[n\theta]    & 0 &       0           &           2\cos^2[n\theta]
\end{array}
\right].
\end{equation}
Here, we assume that the ancillas are initially in the state $\xi^i = \xi^1= 1/2+(\kappa-1/2)\sigma_z$, where $\kappa=0$ corresponds to the pure ground state and $\kappa=1/2$ to the fully mixed state.
We stress that, for this approach to be consistent with our general theory, we need to interpret the norm $\abs{\cdot}$, when applied to a density matrix, as the Euclidean norm of the corresponding vector representation in the operator space of the system. 
That is, if mixed quantum states are treated as operators on the Hilbert space of the system, and thus represented as matrices, $\abs{\cdot}$ must be evaluated as the Hilbert-Schmidt norm. 
In either case, $\nrm{\cdot}$ must be interpreted as the operator norm induced by the norm $\abs{\cdot}$. 
The norms of the maps $\mF_n$ and $\mF_n^{-1}$, for example, are given by 
\begin{equation}
\nrm{\mF_n}= k^n
\quad\text{and} \quad
\nrm{\mF_n^{-1}}= \frac{1}{k^n\cos^2[n\theta]}
\end{equation}
for $\kappa =1/2$. 
For this specific value of $\kappa$, which is considered in the main text, the parameters $v,M$ become 
\begin{equation}
v= (1-k)\cos^2[\theta]=(1-k)(1-u)
\quad\text{and}\quad
M=k(1-k),
\end{equation}
where $u=\sin^2[\theta]$ is the coherent swap probability.

\end{document}